\documentclass[11pt]{article}
\pdfoutput  = 1
\usepackage{latexsym}
\usepackage{amsmath}
\usepackage{verbatim}
\usepackage{times}
\usepackage{amsfonts}
\usepackage{mathrsfs}
\usepackage{eucal}
\usepackage[ruled]{algorithm2e}
\usepackage{url}

\newtheorem{thm}{\textbf{Theorem}}

\newtheorem{lem}[thm]{\textbf{Lemma}}
\newtheorem{lemma}[thm]{\textbf{Lemma}}

\newcommand{\mbx}{\mathbf{x}}

\newcommand{\mby}{\mathbf{y}}

\ifx\pdfoutput\undefined
\usepackage{graphicx}
\else
\usepackage[pdftex]{graphicx}
\fi
\usepackage{subfigure}
\usepackage{url}
\usepackage{appendix}
\def\QED{\mbox{\rule[0pt]{1.5ex}{1.5ex}}}
\def\proof{\noindent\hspace{2em}{\itshape Proof: }}
\def\endproof{\hspace*{\fill}~\QED\par\endtrivlist\unskip}

\date{}
\begin{document}
\begin{titlepage}
\title{Optimal Gossip-Based Aggregate Computation}
\author{
Jen-Yeu Chen\thanks{Department of Electrical Engineering, National DongHwa University, ShouFeng, Hualien 97401, Taiwan, ROC.
E-mail: {\tt jenyeuchen@acm.org}. Supported in part by NSC Grant 97-2218-E259-003.}
\and Gopal
Pandurangan\thanks{Division of Mathematical Sciences, Nanyang Technological 
University, Singapore 648477, and Department of Computer Science, Brown University, Providence, RI 02912. E-mail:~{\tt gopalpandurangan@gmail.com}. Supported in part by NSF Award  CCF-0830476.}
}

\maketitle \thispagestyle{empty}


\vspace*{.4in}



\begin{abstract}
Motivated by applications to modern networking technologies,  there has been   interest in designing  efficient gossip-based protocols for computing aggregate functions. While gossip-based protocols provide robustness due to their randomized nature, reducing the message and time complexity of these protocols is also of paramount importance in the context of resource-constrained networks such as sensor and peer-to-peer networks.

We present the first provably almost-optimal  gossip-based algorithms for aggregate computation that are both time optimal and message-optimal. Given a $n$-node  network, our algorithms guarantee that all the nodes can compute the common aggregates (such as Min, Max, Count, Sum, Average, Rank etc.)  of their values in optimal $O(\log n)$ time and using $O(n \log \log n)$ messages.
Our result improves on the  algorithm  of Kempe et al.~\cite{kempe} that is time-optimal, but uses $O(n \log n)$ messages as well as on the algorithm of Kashyap et al.~\cite{efficient-gossip} that uses $O(n \log \log n)$ messages, but is not time-optimal (takes $O(\log n \log \log n)$ time). Furthermore, we  show that our algorithms can be used to
improve gossip-based aggregate computation in  sparse communication networks, such as in peer-to-peer networks. 

The main technical ingredient of our algorithm is a  technique called {\em distributed random ranking (DRR)} that can be useful in other applications as well.  DRR  gives an efficient distributed procedure to  partition the network into a forest of  (disjoint) trees of small size. Since the  size of each tree is small,  aggregates within each  tree can be efficiently obtained at their respective roots.
All the roots then perform a uniform gossip algorithm on their local aggregates to reach a distributed consensus on the global aggregates.

Our algorithms are non-address oblivious. In contrast, we show a  lower bound of $\Omega(n\log n)$ on the message complexity of any address-oblivious algorithm for computing aggregates. This shows that  non-address oblivious algorithms  are needed to obtain significantly better message complexity. Our lower bound holds regardless of the number of rounds taken or the size of the messages used.   Our lower bound is the first non-trivial lower bound for gossip-based aggregate computation and  also gives the first formal proof that  computing aggregates is strictly harder than
 rumor spreading in the address-oblivious model. 
\end{abstract}

\noindent {\bf Keywords:} Gossip-based protocols, aggregate computation, distributed  randomized protocols, probabilistic analysis, lower bounds.

\end{titlepage}

\vspace*{-0.15in}
\section{Introduction}
\vspace*{-0.1in}
\subsection{Background and Previous Work}
\vspace*{-0.05in}
Aggregate statistics (e.g., Average, Max/Min, Sum, and, Count etc.)  are significantly useful for many applications in networks~\cite{DRG,sparse-aggregation,Jelasity-tocs,kempe,impact-aggregate,aggregation-service,sensys04-1}. These statistics
have to be computed over data stored at individual nodes. For example, in a peer-to-peer network, the average number of files stored at each node or the maximum size of files exchanged between nodes is  an important statistic needed by system designers for optimizing overall performance~\cite{pastry,chord}. Similarly, in sensor networks,  knowing the average or maximum remaining battery power among the sensor nodes is a critical statistic.
Many research efforts have been dedicated to developing scalable and distributed algorithms  for aggregate computation.  Among them
gossip-based algorithms~\cite{S-Boyd-gossip,DRG,geographic-gossip,efficient-gossip,kempe,smart-gossip,separable-gossip,sensys04-2,Gossip-Roberto-podc08,hierarchical-gossip} have recently received significant attention because of their simplicity of  implementation, scalability to large network size, and robustness to frequent network topology changes.
 In a gossip-based algorithm,  each node  exchanges  information with a randomly chosen communication partner in each round.  The randomness inherent in the gossip-based protocols naturally provides robustness, simplicity, and scalability~\cite{Karp-rumor,efficient-gossip}. We refer to~\cite{Karp-rumor,efficient-gossip,kempe} for a detailed discussion on the   advantages of  gossip-based computation over centralized
 and deterministic approaches and their attractiveness to emerging networking technologies such as peer-to-peer, wireless, and sensor networks. This paper focuses on designing efficient gossip-based protocols for aggregate computation that
 have low message and time complexity. This is especially useful in the context of resource-constrained networks
 such as sensor and  wireless  networks, where reducing message and time complexity can yield significant benefits
 in terms of lowering congestion and lengthening node lifetimes.

Much of the early work on gossip  focused on using randomized communication for rumor propagation \cite{demers,Karp-rumor,pittel}. In particular, Karp et al. \cite{Karp-rumor} gave a rumor spreading algorithm (for spreading a single message throughout a network of $n$ nodes)  that takes $O(\log n)$ communication rounds and $O(n \log \log n)$ messages.
It is easy to establish that $\Omega(\log n)$ rounds are needed by any gossip-based rumor spreading algorithm
(this bound also holds for gossip-based aggregate computation).
They also showed that any rumor spreading algorithm needs  at least $\Omega(n \log \log n)$ messages for a class of randomized gossip-based algorithms referred to as {\em address-oblivious} algorithms \cite{Karp-rumor}.    Informally, an algorithm is called address-oblivious  if  the decision to send a message to its communication partner in a round does not depend on  the partner's address. Karp et al.'s algorithm is address-oblivious. For non-address oblivious algorithms, they show a lower bound of $\omega(n)$ messages,
if the algorithm is allowed only $O(\log n)$ rounds.

  Kempe et al. \cite{kempe} were the first to present randomized gossip-based algorithms for computing aggregates. They analyzed a  gossip-based protocol for computing sums, averages, quantiles, and other
aggregate functions. In their scheme for estimating  average, each node selects another random node to which it sends half of its value; a node on receiving a set of values just adds them to its own halved value. Their protocol  takes $O(\log n)$  rounds and uses $O(n \log n)$ messages to converge to the true average in a $n$-node  network. Their protocol is address-oblivious.  The work of Kashyap et al. \cite{efficient-gossip} was  the first to address the issue of reducing the message complexity of gossip-based aggregate protocols, even at the cost of increasing the time complexity.  They presented an algorithm that significantly improves over the message
complexity of the protocol of Kempe et al. Their algorithm uses only $O(n \log \log n)$ messages, but is not time optimal --- it runs in $O(\log n \log \log n)$ time. Their algorithm achieves this $O(\log n/\log \log n)$ factor reduction
in the number of messages by randomly clustering nodes into groups of size $O(\log n)$, selecting representative for each group, and then having the group representatives gossip among themselves. Their algorithm is not  address-oblivious.
For other related work on gossip-based protocols, we refer to \cite{efficient-gossip,DRG} and the references therein.

\vspace*{-0.15in}
\subsection{Our Contributions}
\vspace*{-0.05in}
\begin{table*}
\caption{DRR-gossip vs. other gossip-based algorithms.}\label{table-performance-DRR-gossip}
\begin{center}
\begin{tabular}{lccc}
  \hline \hline
  Algorithm & time complexity & message complexity & address oblivious? \\
  \hline
  efficient gossip~\cite{efficient-gossip} & $O(\log n \log\log n )$ & $O(n\log\log n)$ & no\\ 
  uniform gossip~\cite{kempe}& $O(\log n)$ &$O(n\log n)$ & yes\\
 DRR-gossip~[this paper] & $O(\log n)$ & $O(n\log\log n)$ & no \\

  \hline \hline
 \end{tabular}
\end{center}
\end{table*}
In this paper,  we present  the first provably almost-optimal gossip-based algorithms  for  computing various  aggregate functions that improves upon previous results. Given a $n$-node  network, our algorithms guarantee that all the nodes can compute the common aggregates (such as Min, Max, Count, Sum, Average, Rank etc.)  of their values in
optimal $O(\log n)$ time and using $O(n \log \log n)$ messages.
Our result  (cf. Table~\ref{table-performance-DRR-gossip}) improves on the  algorithm  of Kempe et al. \cite{kempe} that is time-optimal, but uses $O(n \log n)$ messages as well as on the algorithm of Kashyap et al. \cite{efficient-gossip} that uses $O(n \log \log n)$ messages,
but is not time-optimal (takes $O(\log n \log \log n)$ time).

Our algorithms use a simple  scheme called {\em distributed random ranking (DRR)}  that gives an efficient distributed protocol to  partition the network into a forest of  disjoint trees of $O(\log n)$ size. Since the size of each tree is small,  aggregates within each  tree can be efficiently obtained at their respective roots.
All the roots then perform a uniform gossip algorithm on their local (tree) aggregates to reach a distributed consensus on the global aggregates.  Our  idea of forming trees and then doing gossip among the roots of the trees  is similar to the idea  of Kashyap et al.  The main novelty  is that our DRR technique gives a simple and efficient  distributed way of decomposing the network into disjoint trees (groups) which takes
only $O(\log n)$ rounds and $O(n \log \log n)$ messages.  This leads to a simpler and faster algorithm than that
of \cite{efficient-gossip}.  The paper  of \cite{Gossip-Roberto-podc08}  proposes the following heuristic: divide the network into clusters  (called the ``bootstrap phase"), aggregate the data within the clusters --- these are aggregated
in a small subset of nodes  within each cluster called clusterheads; the clusterheads then use gossip algorithm of Kempe et al  to do inter-cluster aggregation; and, finally the clusterheads will disseminate the information to all the nodes in the respective clusters.  It is not clear in~\cite{Gossip-Roberto-podc08} how to efficiently implement  the bootstrap phase of dividing the network into clusters. Also, only numerical simulation results are presented in~\cite{Gossip-Roberto-podc08} to show that their approach gives better complexity than the algorithm of Kempe et al.
It is mentioned {\em without proof} that their approach can take $O(n \log \log n)$ messages and $O(\log n)$ time.
Hence, to the best of our knowledge, our work presents the first rigorous protocol that provably shows these bounds.

Our second contribution is analyzing gossip-based aggregate computation
in sparse  networks.
In  sparse topologies  such as P2P networks, point-to-point communication  between all pairs of nodes (as assumed in gossip-based protocols)
may not be a reasonable assumption. On the other hand, a small
number of neighbors  in such networks makes it feasible to send one message simultaneously to
all  neighbors in one round: in fact, this is a standard assumption in the distributed message passing model \cite{peleg}.  We show how our DRR technique leads to
improved  gossip-based aggregate computation in such (arbitrary) sparse networks, e.g.,
 P2P network topologies such as Chord \cite{chord}.  The improvement
relies on a key property of the DRR scheme that we prove:  {\em height} of each tree produced by DRR in any {\em arbitrary} graph is
   bounded by $O(\log n)$ whp.  In Chord, for example, we show that DRR-gossip takes $O(\log^2 n)$ time whp and $O(n\log n)$ messages. In contrast,  uniform gossip gives $O(\log^2 n)$ rounds and $O(n \log^2 n)$ messages.

Our algorithm is non-address oblivious, i.e., some steps use addresses to decide which partner to communicate in a round. The time complexity of our algorithm is optimal and the message complexity is  within
a factor $o(\log \log n)$ of the optimal. This is because, Karp et al \cite{Karp-rumor} showed a lower bound of $\omega(n)$ for any non-address oblivious
rumor spreading algorithm that operates in $O(\log n)$ rounds.  (Computing aggregates is at least as hard as rumor spreading.)

 Our third contribution is a non-trivial lower bound of $\Omega(n\log n)$ on the message complexity of any address-oblivious algorithm for computing aggregates. This lower bound holds regardless of the number of rounds taken or the size of the messages (i.e., even assuming that nodes that can send arbitrarily long messages).   Our result shows that non-address oblivious algorithms (such as ours) are needed to obtain a significant improvement in message complexity.
 We note that this bound is significantly larger than the $\Omega(n \log \log n)$ messages shown by Karp et al. for rumor spreading. Thus our result  also gives the first formal proof that  computing aggregates is strictly harder than
 rumor spreading in the address-oblivious model. Another implication of our result  is that the algorithm of Kempe et al. \cite{kempe} is asymptotically message optimal for the address-oblivious model.

Our  algorithm, henceforth called {\em DRR-gossip},  proceeds in phases. In phase one, every  node runs the DRR scheme to construct a forest of (disjoint) trees.
In phase two, each tree computes its  local aggregate (e.g., sum or maximum)  by a convergecast process; the  local aggregate is obtained at the root.
Finally in phase three, all the roots  utilize a suitably modified version of the uniform gossip algorithm of Kempe et al.~\cite{kempe} to obtain the global aggregate. Finally, if necessary, the roots forward the global aggregate  to other nodes in their trees.


\vspace*{-0.1in}
\subsection{Organization}
\vspace*{-0.05in}
The rest of this paper is organized as follows. The network model is described in Section~\ref{sec:network model} followed by sections where each phase of the DRR-gossip algorithm is introduced and analyzed separately. The whole DRR-gossip algorithm is summarized in Section~\ref{sec:whole algorithm}. Section~\ref{sec:p2p} applies DRR-gossip to sparse networks. An lower bound on the message complexity of any address-oblivious algorithm for computing aggregates is presented and proved in Section~\ref{sec:lbound-add-oblivious}. 
Section~\ref{sec:prelim}  lists the main probabilistic tools used in our analysis --- the Doob martingale and Azuma's inequality. 
Section \ref{sec:conclusion} concludes with some open questions.

\subsection{Probabilistic Preliminaries}
\label{sec:prelim}

 We  use Doob martingales  extensively in our analysis \cite{random-alg-book}. Let $X_0, \dots, X_n$ be {\em any} sequence of random variables
and let $Y$ be any random variable with $E[|Y|] < \infty$.
Define the random variable $Z_i = E[Y|X_0, \dots, X_i]$, $ i =
0,1,\dots,n$.
Then $Z_0, Z_1, \dots, Z_n$ form a {\bf Doob martingale} sequence.

We use the martingale inequality  known as Azuma's inequality, stated as follows \cite{random-alg-book}.
Let $X_0, X_1, \dots $ be a martingale sequence such that for each
$k$,
\[ |X_k - X_{k-1}| \leq c_k\] where $c_k$ may depend on $k$.
Then  for all $t \geq 0$ and any $\lambda > 0$,
\begin{equation}
\label{eq:azuma} \Pr(|X_t - X_0| \geq \lambda ) \leq 2 e^{-
\frac{\lambda^2}{2\sum_{k=1}^tc_k^2}}
\end{equation}

We also need the following variant of the
Chernoff bound from \cite{Srinivasan}, that works in the case of
dependent  indicator random variables that are  correlated as
defined below.

\begin{lemma}(\cite{Srinivasan})
\label{lem_fkg} Let $Z_1,Z_2,\ldots,Z_s \in \{0,1\}$ be random
variables such that for all $l$, and for any $S_{l-1} \subseteq
\{1,\ldots,l-1\}, \Pr(Z_l = 1 | \bigwedge_{j \in S_{l-1}} Z_j=1)
\leq \Pr(Z_l = 1)$. Then for any $\delta > 0$,
$\Pr(\sum\limits_{l=1}^{s}Z_l \geq \mu(1+\delta)) \leq
(\frac{e^{\delta}}{(1+\delta)^{1+\delta}})^{\mu}$, where $\mu =
\sum\limits_{l=1}^{s}E[Z_l]$.
\end{lemma}
%
%



\vspace*{-0.15in}
\section{Model}\label{sec:network model}
\vspace*{-0.1in}
The network consists of a set $V$ of $n$ nodes; each node $i \in V$ has a data value denoted by $v_i$.
  The goal is to compute aggregate functions such as Min, Max, Sum, Average etc., of the node values.

The nodes communicate in discrete time-steps referred to as {\em rounds}. As in prior works on this problem \cite{Karp-rumor, efficient-gossip}, we assume that communication rounds are synchronized, and all nodes can communicate simultaneously in a given round.   Each node can communicate with every other node. In a round, each node can choose a communication partner independently and uniformly at random.  A node $i$ is said to {\em call } a node $j$ if $i$ chooses $j$ as a communication partner.  (This is known as the {\em random phone call} model \cite{Karp-rumor}.) Once a call is established, we assume that information can be exchanged in both directions along the link.   In one round, a node can call only {\em one}  other node.   We assume that nodes have unique addresses.
 The length of a message is limited to $O(\log n + \log s)$, where $s$ is the range of values.
 It is important to limit the size of messages  used in aggregate computation, as communication bandwidth is often a costly resource in distributed settings.  All the above assumptions
are also used in prior works \cite{efficient-gossip,kempe}.
Similar to the algorithms of  \cite{efficient-gossip,kempe}, our algorithm can tolerate the following two types of failures: (i) some fraction of nodes may crash initially, and (ii) links are lossy and messages can get lost.
Thus, while nodes cannot fail once the algorithm has started, communication can fail with a certain probability
$\delta$.
Without loss of generality, $1/\log n < \delta < 1/8$:  Larger values of $\delta$, requires only $O(1/\log(1/\delta))$ repeated calls to bring down the probability below $1/8$,  and smaller values only make it easier to prove our claims.

Throughout the paper, ``with high probability (whp)" means ``with probability at least $1 - 1/n^{\alpha}$, for some $\alpha > 0$".

\begin{algorithm}[t]
\label{alg:DRR}
  \caption{$\mathbb{F}=$DRR($G$)}
  \ForEach{node $i\in V$}
  {
  choose $rank(i)$ independently and uniformly at random from $[0,1]$ \;
  set $found$ = FALSE   // higher ranked node not yet found \;
  set $parent(i)=\text{NULL}$ // initially every node is a root node\;
  set $k = 0$   // number of random nodes probed \;
  \Repeat{$found ==TRUE$ or $k < \log n - 1$}
  {sample  a node  $u$  independently and uniformly at random from $V$ and get its rank \;
   \If{$rank(u)  >  rank(i)$}
   {set $parent(i) = u$\; set $found = TRUE$\;
   set $k = k+ 1$\;}
   }

   \If{$found == TRUE$}
   {send a connection message including its identifier, $i$, to its parent node $parent(i)$\;}
   Collect the connection messages and accordingly construct the set of its children nodes, $Child(i)$\; 
   \eIf{$Child(i)=\emptyset$}
   {become a leaf node\;}
   {become an intermediate node\;}
   } 
\end{algorithm}

\vspace*{-0.15in}
\section{DRR-Gossip Algorithms}
\vspace*{-0.1in}
\subsection{Phase I: Distributed Random Ranking (DRR)}
\vspace*{-0.1in}
The DRR algorithm is as follows (cf. Algorithm~\ref{alg:DRR}).
Every node $i\in V$ chooses  a rank independently and uniformly
at random from $[0,1]$. (Equivalently, each node can  choose a  rank  uniformly at random from $[1,n^3]$ which leads to the same asymptotic bounds; however, choosing from $[0,1]$ leads to a smoother analysis, e.g., allows use of integrals.)  Each node $i$  then
samples up to $\log n -1$   random nodes sequentially (one in each round) till it finds a node of higher rank to connect to.  If none of the $\log n -1$ sampled  nodes have a higher rank then node $i$ becomes a ``root".  Since every node except root nodes connects to a node with higher rank, there is no cycle in the graph.   Thus this process results in a collection of disjoint trees which together constitute a forest $\mathbb{F}$.  

In the following two theorems, we show the upper bounds of the number of trees and  the size of each tree produced by the DRR algorithm; these are critical in bounding the time complexity of DRR-gossip.


\begin{thm}[Number of Trees]
\label{le:numtrees}
The number of  trees produced by the DRR algorithm is $O(n/\log n)$ whp.
\end{thm}
\vspace*{-0.1in}
\proof
%
%
Assume that ranks have already been assigned to the nodes.  All ranks are distinct with probability~1.   Number the nodes according to the order statistic of their ranks: the $i$th node is the node with
the $i$th smallest rank. Let the indicator random variable $X_i$  take the value of 1 if the $i$th smallest node is
a root and 0 otherwise.  
Let $X = \sum_{i=1}^n X_i$ be the total number of roots.  
The $i$th smallest node becomes a root if all the nodes that it samples have rank smaller than or equal to itself, i.e.,
 $ \Pr(X_i = 1) =  \left( \frac{i}{n}\right)^{\log n - 1}.$
 Hence, by linearity of expectation, the expected number of roots (and thus, trees) is:
 $$E[X] = \sum_{i=1}^n \Pr \left(X_i = 1\right) = \sum_{i=1}^n \left( \frac{i}{n}\right)^{\log n - 1} = \Theta \left( \int_1^n \left(\frac{i}{n}\right)^{\log n - 1}\,d\,i \right) =   \Theta \left( \frac{n}{\log n} \right).$$
Note that $X_i$s are independent (but  not identically distributed) random variables, since the probability that the $i$th smallest ranked
node becomes the root depends only on the $\log n-1$  random nodes that it samples and independent of the samples of the rest of the  nodes.  Thus, applying a  Chernoff's bound \cite{random-alg-book}, we have:

$\Pr(X > 6 E[X]) \leq 2^{E[X]} = o(1/n).$
\endproof



\vspace*{-0.1in}
\begin{thm}[Size of a tree]
\label{le:sizeoftree}
The number of nodes in every tree produced by the DRR algorithm is at most $O(\log n)$ whp.
\end{thm}
\vspace*{-0.1in}
\proof
We  bound that the probability that a tree of size $\Omega(\log n)$ is produced by the DRR algorithm.
Fix a set $S$ of $k = c \log n$ nodes, for some sufficiently large positive constant $c$.
We first compute the probability that this set of $k$ nodes form a tree.
For the sake of analysis, we will direct tree edges as follows: a tree edge $(i,j)$ is directed  from  node $i$ to node $j$ if  $rank(i) < rank(j)$, i.e. $i$ connects to $j$. Without loss of generality, fix a permutation of $S$: ($s_1, \ldots, s_{\alpha}, \ldots,s_{\beta}, \ldots, s_k)$ where $rank(s_{\alpha})>rank(s_{\beta}), \quad 1\le \alpha < \beta \le k$. This permutation  induces
 a  directed spanning tree  on $S$ in the following sense: $s_1$ is the root and  any other node $s_\alpha$ ($1 < \alpha \le k$) connects to a node in the totally (strictly) ordered set $\{s_1, \dots, s_{\alpha-1}\}$ (as fixed by the above permutation).
 For convenience, we denote the event that a node $s$ connects to any node on a directed  tree, $T$, as $s\rightarrow T$. Note that $s\rightarrow T$ implies that $s$'s rank is less than that of any node on the tree $T$. Also, we denote  the event of a directed spanning tree being induced on the totally (strictly) ordered set $\{s_1,s_2,\dots, s_{\alpha},\ldots,s_h\}$ as $T_h$, where a node $s_{\alpha}$ can only connect to its preceding nodes in the ordered set. As a special case, $T_1$ is the event of the induced directed tree containing only the root node $s_1$.  We are interested in the event $T_k$, i.e.,  the
 set $S$ of $k$ nodes forming a directed spanning tree in the above fashion. 
 %
 %
 In the following, we bound the probability of  the event $T_k$ happening:
 \begin{align}
 \Pr(T_k)&=\Pr\left(T_1 \cap (s_2 \rightarrow T_1) \cap (s_3 \rightarrow T_2)  \cap \dots \cap (s_k \rightarrow T_{k-1})\right) \nonumber\\
 &= \Pr(T_1)  \Pr(s_2 \rightarrow T_1 |\; T_1) \Pr(s_3 \rightarrow T_2 |\, T_2) \dots \Pr(s_k \rightarrow T_{k-1}|\;  T_{k-1}).
 \label{eq:prob_of_tree}
\end{align}
 To bound each of the terms in the product, we  use the principle of deferred decisions: when a new node is sampled (i.e., for the first time) we assign it a random rank. For simplicity, we assume that each node sampled is a new node --- this does not change the asymptotic bound, since there are now only $k = O(\log n)$ nodes under consideration and each node samples at most $O(\log n)$  nodes. This assumption allows us to use the principle of deferred decisions to assign random ranks  without worrying about sampling an already sampled node. Below we bound the conditional probability $\Pr( s_{\alpha}\rightarrow T_{\alpha-1} |\; T_{\alpha-1})$, for any $2 \leq \alpha \leq k$ as follows. Let $r_{q}=rank(s_{q})$ be the rank of node $s_{q}$, $1\le q \le \alpha$; then
$$\Pr( s_{\alpha}\rightarrow T_{\alpha-1} |\; T_{\alpha-1})
 \leq  \int_0^ 1 \int_0^{r_1}\int_0^{r_2} \dots \int_0^{r_{\alpha -1}}\; \sum_{h=0}^{\log n - 1} \left(\frac{\alpha -1}{n}\right)r^h_{\alpha} \; dr_{\alpha} \dots dr_1.$$
 The explanation for the above bound is as follows: Since $T_{\alpha -1}$ is a directed spanning tree on the first
 $\alpha -1$ nodes, and $s_{\alpha}$ connects to $T_{\alpha - 1}$, we have $r_1 > r_2 > \dots > r_{\alpha - 1} > r_{\alpha}$.  Hence $r_1$ can take any value between 0 and 1,
 $r_2 $ can take any value between 0 and $r_1$ and so on. This is captured by the respective ranges of the integrals.
 The term inside the integrals is explained as follows.
 There  are at most $\log n- 1$ attempts for node $s_{\alpha}$  to connect to any one of the first $\alpha - 1$ nodes. Suppose, it connects in the $h$th attempt. Then, the first
 $h-1$ attempts should connect to nodes whose rank should be less than $r_{\alpha}$, hence the term $r_{\alpha}^h$ (as mentioned earlier, we assume that we don't sample an already sampled node, this doesn't change the bound asymptotically).
 The term $(\alpha -1)/n$ is the probability that $s_{\alpha}$ connects to any one of the first $\alpha - 1$ nodes in the $h$th attempt.

Simplifying the right hand side, we have,
 \begin{align*}
 \Pr&( s_{\alpha}\rightarrow T_{\alpha-1} |\; T_{\alpha-1}) \\
 & \leq \frac{{\alpha}-1}{n} \int_0^ 1 \int_0^{r_1}\int_0^{r_2} \dots \int_0^{r_{{\alpha}-1}} [1 + r_{\alpha} + r_{\alpha}^2 + \dots r_{\alpha}^{\log n - 1}]  dr_{\alpha} \dots dr_1  \\
&= \frac{{\alpha}-1}{n} \left( \frac{0!}{{\alpha}!} +  \frac{1!}{({\alpha}+1)!} + \frac{2!}{({\alpha}+2)!} + \dots + \frac{(\log n)!}{(\log n + {\alpha})!}\right).
\end{align*}

The above expression is bounded by
$\frac{b}{n}$,
where $0< b <1$ if ${\alpha} > 2$ and $0< b \leq (1 - \frac{1}{\log n + 2})$ if $\alpha = 2$.
Besides, $\Pr(T_1) \leq \frac{1}{\log n}$ (cf. Theorem~\ref{le:numtrees});
hence, the equation~(\ref{eq:prob_of_tree}) is bounded by
$\left(\frac{b}{n} \right)^{k-1} \frac{1}{\log n}.$

Using the above, the probability that a tree of size $k = c\log n$ is produced by the DRR algorithm is bounded by
\[
{n \choose k} k! (\frac{b}{n})^{k-1}\frac{1}{\log n}
\leq \frac{(ne)^k}{k^k} O(\sqrt{k}) \frac{k^k}{e^k} (\frac{b}{n})^{k-1}\frac{1}{\log n}
 \leq \frac{c' \cdot n}{\log ^{\frac{1}{2}}n} \cdot b^{k-1}
=o(1/n),\]
if  $c$ sufficiently large.
\endproof

\noindent {\bf Complexity of Phase I --- the DRR algorithm}
\vspace*{-0.05in}
\begin{thm}
The message complexity of the DRR algorithm is $O(n \log \log n)$ whp. The time complexity is $O(\log n)$ rounds.
\end{thm}
\vspace*{-0.1in}
\proof
Let $d = \log n -1$.
Fix a node $i$. Its
 rank is chosen uniformly at random from $[0,1]$.
The expected number of nodes sampled before a node $i$ finds a higher ranked node (or else,  all $d$ nodes will be sampled)
is computed as follows.
The probability that exactly $k$ nodes will be sampled is $\Theta(\frac{1}{k+1}\frac{1}{k})$, since
the last node sampled should be the highest ranked node and $i$ should be the second highest ranked node (whp, all the nodes sampled will be unique).
Hence the expected number of nodes  probed  is
$ \sum_{k = 1}^{d} \Theta\left(k \frac{1}{k+1}\frac{1}{k}\right) = O(\log d).$
Hence the number of messages exchanged by node $i$ is $O(\log d)$.
By linearity of expectation, the total number of messages exchanged by all nodes
is $O(n \log d) = O(n \log \log n)$.

To show concentration, we set up a Doob martingale as follows. Let $X$ denote the random variable that counts the
total number of nodes sampled by all nodes. $E[X] = O(n \log d)$. Assume that ranks have already been assigned to the nodes. Number the nodes according to the order statistic of their ranks: the $i$th node is the node with
the $i$th smallest rank.
Let the indicator r.v.  $Z_{ik}$  ($1 \leq i \leq n$, $1 \leq k \leq d$)  indicate whether the $k$th sample by the $i$th {\em smallest ranked}  node succeeded or not (i.e., it found
a higher ranked node).  If it succeeded then $Z_{ij} = 1$ for all $j \leq k$ and $Z_{ij} = 0$ for all $j > k$.
Thus $X = \sum_{i = 1}^n \sum_{k =1}^d Z_{ik}$.
Then  the sequence $X_0 = E[X], X_1 = E[X|Z_{11}], \dots,  X_{nd} = E[X|Z_{11}, \dots, Z_{nd}]$ is a Doob martingale. Note that $|X_{\ell} - X_{\ell-1}| \leq d$ ($1 \leq \ell \leq nd$) because fixing the outcome of a sample of one node affects only
the outcomes of other samples made by the same node and not the samples made by other nodes.
Applying Azuma's inequality, for a positive constant $\epsilon$ we have:
$$\Pr(|X-E[X]| \geq \epsilon n) \leq 2 \exp\left(-\frac{\epsilon^2 n^2}{2 n (\log n)^3}\right) = o(1/n).$$

The time complexity is immediate since each node probes at most $O(\log n)$ nodes in as many rounds.
\endproof




\vspace*{-0.15in}
\subsection{Phase II: Convergecast and Broadcast}
\vspace*{-0.1in}
 In the second phase of our algorithm, the local aggregate of each tree is obtained at the root by the Convergecast algorithm --- an aggregation process starting from leaf nodes and proceeding upward along the tree to the root node. For example, to compute the local max/min, all leave nodes simply send their values to their parent nodes. An intermediate node collects the values from its children, compares them with its own value and sends its parent node the max/min value among all received values and its own. A root node then can obtain the local max/min value of its  tree. Algorithm~\ref{alg:convergecast-max} and  Algorithm~\ref{alg:convergecast-sum}
  are the
 pseudo-codes of the Convergecast-max algorithm and the Convergecast-sum algorithm, respectively. 

After the Convergecast process, each root broadcasts its address to all other nodes in its tree via the tree links. This process proceeds from the root down to the leaves via the tree links (these two-way links were already established during Phase 1.)
At the end of this process, all non-root nodes  know the identity (address) of their respective roots.

\begin{algorithm}[t]
\label{alg:convergecast-max}
\caption{$\mathbf{cov}_{max}=$convergecast-max($\mathbb{F}$,$\mathbf{v}$)}
\KwIn{the ranking forest $\mathbb{F}$, and the value vector $\mathbf{v}$ over all nodes in $\mathbb{F}$}
\KwOut{the local $Max$ aggregate vector $\mathbf{cov}_{max}$ over roots}
\lForEach{leaf node}{send its value to its parent\;}
\ForEach{intermediate node}
 {
  - collect values from its children\;
  - compare collected values with its own value\;
  - update its value to the maximum amid all and send the maximum to its parent.
 }
\ForEach{root node $z$}
 {
 - collect values from its children\;
 - compare collected values with its own value\;
 - update its value to the local maximum value $\mathbf{cov}_{max}(z)$.
}
\end{algorithm}

\begin{algorithm}[h!]
\label{alg:convergecast-sum}
\caption{$\mathbf{cov}_{sum}=$convergecast-sum($\mathbb{F}$,$\mathbf{v}$)}
\KwIn{
the ranking forest $\mathbb{F}$ and the value vector $\mathbf{v}$ over all nodes in $\mathbb{F}$}
\KwOut{the local $Ave$ aggregate vector $\mathbf{cov}_{max}$ over roots.}
{\bf Initialization}: every node $i$ stores a row vector $(v_i, w_i=1)$ including its value $v_i$ and a size count $w_i$\;
\ForEach{leaf node $i \in \mathbb{F}$}
{- send its parent a message containing the vector $(v_i, w_i=1)$\;
 - reset $(v_i, w_i)=(0,\,0)$.}
\ForEach{intermediate node $j \in \mathbb{F}$}{
 - collect messages (vectors) from its children\;
 - compute and update $v_j=v_j+\sum_{k\in Child(j)}v_k,$ and $w_j=w_j+\sum_{k\in Child(j)}w_k,$ where $Child(j)=\{\text{$j$'s children nodes}\}$\;
 - send computed $(v_j, w_j)$ to its parent\;
 - reset its vector $(v_j, w_j)=(0,\,0)$ when its parent successfully receives its message.
 }
\ForEach{root node $z \in \tilde{V}$}{
 - collect messages (vectors) from its children\;
 - compute the local sum aggregate
$\mathbf{cov}_{sum}(z,1)=v_z+\sum_{k\in Child(z)}v_k,$ and the size count of the tree $\mathbf{cov}_{sum}(z,2)=w_z+\sum_{k\in Child(z)}w_k,$ where $Child(z)=\{\text{$z$'s children nodes}\}$.
}
\end{algorithm}

\noindent {\bf Complexity of Phase II}

Every node except the root nodes needs to send a message to its parent in the upward aggregation process of the Convergecast algorithms. So the message complexity  is $O(n)$.
 Since each node can communicate with at most one node in one round,
 the time  complexity is bounded by the size of the tree.
(This is the reason for bounding size and not just the height.)
 Since the  tree size (hence, tree height also) is bounded by $O(\log n)$ (cf. Theorem~\ref{le:sizeoftree}) the time complexity of Convergecast and Broadcast is $O(\log n)$. Moreover, as the number of roots is at most $O(n/\log n)$ by Theorem~\ref{le:numtrees}, the message complexity for broadcast is also $O(n)$.

\vspace*{-0.15in}
\subsection{Phase III: Gossip}
\vspace*{-0.1in}
In the third phase, all roots of the trees compute the global aggregate by performing the uniform gossip algorithm on the graph $\tilde{G}=clique(\tilde{V})$, where $\tilde{V}\subseteq V$ is the set of roots and $|\tilde{V}|=m = O(n/\log n)$.

The idea of uniform gossip is as follows. Every root independently and uniformly at random selects a node to send  its message. If the selected node is another root then the task is completed. If not,  the selected node needs to forward the received message to its root (all nodes in a tree know the root's address at the end of Phase II ---  here is where we use a
non-address oblivious communication). Thus, to traverse through an edge of $\tilde{G}$, a message needs at most two hops of $G$.

 Algorithm~\ref{alg:gossip-max}, Gossip-max, and  Algorithm~\ref{alg:gossip-ave}, Gossip-ave (which is a modification from the  Push-Sum algorithm of~\cite{efficient-gossip,kempe})  compute the $Max$ and $Ave$ aggregates respectively (other aggregates such as Min, Sum etc., can be calculated by a suitable modification). Note that,  unlike  Gossip-max,  Gossip-ave algorithm does not need a  sampling procedure.

Algorithm~\ref{alg:data-spread}, Data-spread, a modification of  Gossip-max,  can be  used by  a root node to spread its value. If a root needs to spread a particular value over the network, it sets this  value as its initial value  and all  other roots set their initial value to minus infinity.


%
%
%
\begin{algorithm}[h!]
\label{alg:gossip-max}
\caption{ $\mathbf{\hat{\mbx}}_{max}=$Gossip-max($G,\,\mathbb{F},\,\tilde{V},\,\mathbf{y}$)}
{\bf Initialization}: every root $i\in \tilde{V}$ is of the initial value $x_{0,i}=y(i)$ from the input $\mathbf{y}$.
\\ /* To compute $Max$, $x_{0,i}=y(i)=\mathbf{cov}_{max}(i)$;
  To compute $Ave$, $x_{0,i}=y(i)=\mathbf{cov}_{sum}(i,2). */$\;
{\bf Gossip procedure}:\;
\For{$t$=1 : $O(\log n)$ rounds}{
 Every root $i\in \tilde{V}$ independently and uniformly at random, selects a node in $V$ and sends the selected node a message containing its current value $x_{t-1,i}$.\;
 Every node $j\in V-\tilde{V}$ forwards any received messages to its root.\;
 Every root $i\in \tilde{V}$\;
 \Indp
 --- collects messages and compares the received values with its own value\;
 --- updates its current value $x_{t,i}$, which is also the $\hat{\mbx}_{max,t}(i)$, node $i$'s current estimate of $Max$, to the maximum among all received values and its own.\;
}
{\bf Sampling procedure:}\;
\For{$t$=1 : $\frac{1}{c}\log n$ rounds}{
Every root $i\in \tilde{V}$ independently and uniformly at random selects a node in $V$ and sends each of the selected nodes an inquiry message.\;
Every node $j\in V-\tilde{V}$ forwards any received inquiry messages to its root.\;
Every root $i\in \tilde{V}$, upon receiving inquiry messages, sends the inquiring roots its value.\;
Every root $i\in \tilde{V}$, updates $x_{t,i}$, i.e. $\hat{\mbx}_{max,t}(i)$, to the maximum value it inquires.}
\end{algorithm}
\begin{algorithm}
\label{alg:data-spread}
\caption{  $\hat{\mbx}_{ru}=$Data-spread($G,\,\mathbb{F},\,\tilde{V},\,x_{ru}$)}
{\bf Initialization}: A root node $i\in \tilde{V}$ which intends to spread its value $x_{ru}$, $|x_{ru}|<\infty$ sets $x_{0,i}=x_{ru}$. All the other nodes $j$ set $x_{0,j}= -\infty$.\;
Run gossip-max($G,\,\mathbb{F},\,\tilde{V},\,\mbx_0$) on the initialized values.
\end{algorithm}
\begin{algorithm}[!ht] 
\label{alg:gossip-ave}
\caption{  $\mathbf{\hat{\mbx}}_{ave}=$Gossip-ave($G,\,\mathbb{F},\,\tilde{V},\,\mathbf{cov}_{sum}$)}
{\bf Initialization}: Every root $i\in \tilde{V}$ sets a vector $(s_{0,i}, g_{0,i})=\mathbf{cov}_{sum}(i)$, where $s_{0,i}$ and $g_{0,i}$ are the local sum of values and the size of the tree rooted at $i$, respectively.\;
\For{$t=1\,:\,O(\log m +\log(1/\epsilon))$ rounds}{
Every root node $i\in \tilde{V}$ independently and uniformly at random selects a node in $V$ and sends the selected node a message containing
 a row vector $(s_{t-1,i}/2, g_{t-1,i}/2)$.\;
Every node $j\in V-\tilde{V}$ forwards any received messages to the root of its ranking tree.\;
Let  $A_{t,i}\subseteq \tilde{V}$ be the set of roots whose messages reach root node $i$ at round $t$. Every root node $i\in \tilde{V}$ updates its row vector by\;
\Indp
 $s_{t,i}=s_{t-1,i}/2+\sum_{j\in A_{t,i}}s_{t-1,j}/2$,\;
 $g_{t,i}=g_{t-1,i}/2+\sum_{j\in A_{t,i}} g_{t-1,j}/2$.\;
\Indm
Every root node $i\in \tilde{V}$ updates its estimate of the global average by $\hat{\mbx}_{ave, t}(i)=\hat{x}_{ave, t, i}=s_{t,i}/g_{t,i}$.
}
\end{algorithm}

\vspace*{-0.1in}
\subsubsection{Performance of Gossip-max and Data-spread Algorithms}
\vspace*{-0.05in}
Let $m$ denote the number of  root nodes. By Theorem~\ref{le:numtrees}, we have $m=|\tilde{V}|=O(n/\log n)$ where $n=|V|$. Karp, et al.~\cite{Karp-rumor} show that all  $m$ nodes of a complete graph can know a particular rumor (e.g., the $Max$ in our application) in $O(\log m)=O(\log n)$ rounds with high probability
by using their Push algorithm (a prototype of our Gossip-max algorithm)  with  uniform selection probability. Similar to the Push algorithm,  Gossip-max  needs $O(m\log m)=O(n)$ messages for all  roots to obtain $Max$ if the selection probability is uniform, i.e., $1/m$.
However, in the implementation of the Gossip-max algorithm on the  forest, the root of a tree is selected with a \emph{probability proportional to its size (number of nodes in the tree)}. Hence, the selection probability is not uniform. In this case, we can only guarantee that after  the gossip procedure of the Gossip-max algorithm, a portion of the roots including the root of the largest tree will possess the $Max$. After the gossip procedure, roots can sample  $O(\log n)$ number of other roots to confirm and update, if necessary, their values and reach  consensus on the global maximum, $Max$.
%
%
%
%
%

We show the following theorem for Gossip-Max 
\vspace*{-0.1in}
\begin{thm}\label{thm:uniform}
After the gossip procedure of the Gossip-max algorithm, at least $\Omega(\frac{c\cdot n}{\log n})$ root nodes obtain  the global maximum, $Max$, whp, where $n=|V|$ and $0<c<1$ is a constant. 
\end{thm}
\vspace*{-0.05in}
%
%
%
%
\proof
As per our failure model, a message may fail to reach the selected root node with probability
$\rho$ (which is at most  $2\delta$, since failure may occur either during the initial call to a non-root node or during the forwarding call from the non-root node to the root of its tree).
For convenience, we call those roots who  know the $Max$  value (the global Maximum) as the max-roots and those who do not as the non-max-roots.

Let $R_t$ be the number of max-roots in round $t$.
Our proof is in two steps.
We first show that, whp, $R_t > 4\log n$ after $8\log n/(1-\rho)$ rounds of Gossip-max.
If $R_0 > 4\log n$ then the task is completed. Consider the case when $R_0 < 4\log n$.  Since the initial number of max-roots is small in this case, the chance that a max-root selects another max-root is small. Similarly, the chance that two or more max-roots select the same root is also small.  So, in this step, whp a max-root will select a non-max-root to send out its gossip message. If the gossip message successfully reaches the selected non-max-root, the $R_t$ will increase by 1. Let $X_i$ denote the indicator of the event that a gossip message $i$ from some max-root successfully reaches the selected non-max-root. We have $Pr(X_i=1)=(1-\rho)$. Then $X=\sum_{i=1}^{8\log n/(1-\rho)}X_i$  is the minimal number of max-roots after $8\log n/(1-\rho)$ rounds.  Clearly, $E[X]= 8\log n$. Here we conservatively assume the worst situation that initially there is only one max-root and at each round only one max-root selects a  non-max-root. So $X$ is the minimal number of max-roots after $8\log n/(1-\rho)$ rounds.

Applying Azuma's inequality~\cite{random-alg-book} and setting $\epsilon=1/2$:
\begin{align*}
Pr(|X&-E[X]|>\epsilon E[X]) < 2\exp\left(-\frac{\epsilon ^2E[X]^2}{2(\frac{8\log n}{1-\rho})} \right) \\&< 2\exp\left(-\frac{\frac{1}{4}E[X]^2}{16\log n} \right)
= 2\exp\left(-\log n\right)=2\cdot n^{-1}.
\end{align*}
Hence, with probability at least $1-\frac{2}{n}$, after $8\log n/(1-\rho)=O(\log n)$ rounds, $R_t>\frac{1}{2}E[X]=4\log n$.

In the second step of our proof, we find the \emph{lower bound of the increasing rate} of $R_t$ when $R_t>4\log n$.
In each round, there are $R_t$ messages sent out from max-roots. Let $Y_i$ denote the indicator of an event that such an message $i$ from a max-root successfully reaches a non-max-root. The $Y_i=0$ when one of the following event happens. (1)  The message $i$ fails in routing to its destination in probability $\rho$. (2)  The  message $i$ destined to another max-root although it successfully travels over the network with probability $(1- \rho)$. The probability of this event is at most $\frac{ (1- \rho)R_t\log n}{n}$ since whp the size of a ranking tree is $O(\log n)$ (cf.Theorem~\ref{le:sizeoftree}).  (3) The message $i$ and at least one another message are destined to the same non-max-root. As the probability three or more messages are destined to a same node is very small, we only consider the case that two messages select the same non-max-root. We also conservatively exclude both two messages on their possible contributions to the increase of $R_t$. This event happens with the probability at most  $\frac{(1- \rho)R_t\log n}{n}$.

Applying union bound~\cite{random-alg-book},
\begin{align*}
Pr(Y_i=0)\leq \rho + \frac{2(1- \rho)R_t\log n}{n}.
\end{align*}
Since $R_t\le \frac{c n}{\log n}$ for any constant $0<c<1$ (otherwise, the task is completed),
\begin{align*}
Pr(Y_i=0)\leq \rho +  2c(1- \rho)=c'+(1-c')\rho,
\end{align*}
where $c'=2c<1$ is a constant that is suitably fixed so that $c'+(1-c')\rho <1$. Consequently, we have
\(Pr(Y_i=1)> (1-c')(1-\rho),\)
and
\(E[Y]=\sum_{i=1}^{R_t}E[Y_i] > (1-c')(1-\rho)R_t.\)
Applying Azuma's inequality,
\begin{align*}
Pr(|Y-E[Y]|&>\epsilon E[Y]) < 2\exp\left( -\frac{\epsilon^2 E[Y]^2}{2R_t}\right)\\
&< 2\exp\left( -\frac{\epsilon^2(1-c')^2(1-\rho)^2R_t}{2} \right).
\end{align*}
Since in this step, whp $R_t>4\log n$, and $(1-c')^2(1-\rho)^2>0$, setting $\epsilon =\frac{1}{2}$ and $\alpha = O(1)$, we obtain
\begin{align*}
Pr(Y<\frac{1}{2}(1-c')(1-\rho)R_t) < 2\cdot n^{-\alpha}.
\end{align*}
Thus, whp,
\(R_{t+1} > R_t+ \frac{1}{2}(1-c')(1-\rho)R_t=\beta R_t,\)
where $ \beta=1+ \frac{1}{2}(1-c')(1-\rho) >1$.
Therefore, whp, after $(8\log n/(1-\rho)+\log_{\beta}n)=O(\log n)$ rounds, at least $\Omega(\frac{c\cdot n}{\log n})$ roots will have the $Max$.
\endproof
%
%
%
%

\noindent {\bf Sampling Procedure}

From Theorem~\ref{thm:uniform}, after the gossip procedure, there are $\Omega(\frac{c n}{\log n})=\Omega(c m)$, $0<c<1$ nodes with the $Max$ value.
For   roots to reach  consensus on $Max$,   they sample each other as in the sampling procedure.
It is possible that the root of a larger  tree will be sampled more frequently than the roots of smaller trees.  However, this non-uniformity is an advantage, since the roots of larger trees would have obtained  $Max$ (in the gossip procedure) with higher probability due to this same non-uniformity. Hence, in  the sampling procedure, a root without $Max$ can obtain $Max$ with higher probability by this non-uniform sampling.
Thus, we have the following theorem 
\begin{thm}\label{thm:sampling}
After the sampling procedure of  Gossip-max algorithm, all  roots know the $Max$ value, whp.
\end{thm}
\vspace*{-0.1in}
%
\proof
After the sampling procedure, the probability that none of the roots possessing the $Max$ is sampled by a root not knowing the $Max$ is at most
\(\left(\frac{m - cm}{m}\right)^{\frac{1}{c}\log n}<\frac{1}{n}.\)
Thus, after the sampling procedure, with probability at least $1-\frac{1}{n}$, all the roots will know the $Max$.
\endproof
%

\noindent {\bf Complexity of Gossip-max and   Data-spread algorithms}

The gossip procedure takes $O(\log n)$ rounds and $O(m\log n)$$=$$O(\frac{n}{\log n}\log n)$$=$$O(n)$ messages. The sampling procedure takes $O(\frac{1}{c}\log n)$$=$$O(\log n)$ rounds and $O(\frac{m}{c}\log n)$$=$$O(n)$ messages.
To sum up, this phase totally takes $O(\log n)$ rounds and $O(n)$ messages for all the roots in the network to reach consensus on $Max$.
The complexity of  Data-spread algorithm is the same as  Gossip-max algorithm.
\subsubsection{Performance of  Gossip-ave Algorithm}
When the uniformity assumption holds in gossip (i.e., in each round, nodes are selected uniformly at random),  it has been shown in~\cite{kempe} that on an $m$-clique with probability at least $1-\delta'$, Gossip-ave (uniform push-sum in~\cite{kempe}) needs $O(\log m+\log \frac{1}{\epsilon}+\log \frac{1}{\delta'})$ rounds and $O(m(\log m+\log \frac{1}{\epsilon}+\log \frac{1}{\delta'}))$ messages for all  $m$ nodes to reach consensus on the global average within a relative error of at most $\epsilon$.
When  uniformity  does not hold, the performance of uniform gossip will depend on the distribution of  selection probability. In efficient gossip algorithm ~\cite{efficient-gossip}, it is shown that the node being selected with the largest probability will have the global average, $Ave$, in $O(\log m + \log \frac{1}{\epsilon})$ rounds. Here, we prove  that the same upper bound holds for our Gossip-ave algorithm, namely,  the root of the largest tree  will have $Ave$ after $O(\log m + \log \frac{1}{\epsilon})$ rounds of the gossip procedure of  Gossip-ave algorithm. In this bound, $m=O(n/\log n)$ is the number of roots (obtained from the DRR algorithm) and the relative error $\epsilon = n^{-\alpha },\,\alpha >0$.
\begin{thm}\label{thm:gossip-ave-upper-bound}
Whp, there exists a time $T_{ave}=O(\log m +\alpha \log n)=O(\log n)$, $\alpha >0$, such that for all time $t\ge T_{ave}$, the relative error of the estimate of average aggregate on the root of the largest ranking tree, $z$, is at most $\frac{2}{n^{\alpha}-1}$, where the relative error is $\frac{|\hat{x}_{ave, t, z}-x_{ave}|}{|x_{ave}|}$, and the average aggregate, $Ave$, is $x_{ave}=\frac{\sum_{i} v_i}{n}$. 
\end{thm}
%

We recall that the gossip-ave algorithm works on the graph $\tilde{G}=clique(\tilde{V})$, where $\tilde{V}\subseteq V$ is the set of roots and $|\tilde{V}|= m = O(n/\log n)$.
To prove Theorem~\ref{thm:gossip-ave-upper-bound},
we need some definitions as in \cite{kempe}.  We define a $m$-tuple contribution vector $\mathbf{y}_{t,i}$ such that $s_{t,i}=\mathbf{y}_{t,i}\cdot \mbx=\sum_{j} y_{t,i,j}x_{j}$ and $w_{t,i}=\|\mby_{t,i}\|_1=\sum_j y_{t,i,j}$,
where $y_{t,i,j}$ is the $j$-th entry of $\mathbf{y}_{t,i}$ and $x_j$ is the initial value at root node $j$, i.e., $x_j=\mathbf{cov}_{sum}(j)$ , the local aggregate of the tree rooted at node $j$ computed by Convergecast-sum.  $\mby_{0,i}=e_i$, the unit vector with the $i$-th entry being  1. Therefore, $\sum_{i} y_{t,i,j}=1$, and $\sum_{ i} w_{t,i}=m$. When $\mby_{t,i}$ is close to $\frac{1}{m}\mathbf{1}$,  where $\mathbf{1}$ is the vector with all entries 1, the approximate of $Ave$, $\hat{x}_{ave, t, i}=\frac{s_{t,i}}{g_{t,i}}$, is close to the true average $x_{ave}$. Note that $w_{t,i}$, which is different from $g_{t,i}$, is a dummy parameter borrowed from~\cite{kempe} to characterize the diffusion speed.

In our Gossip-ave algorithm, we set $g_{0,i}$ to be the size of the root $i$'s  tree. The algorithm then computes the estimate of average directly by $\hat{x}_{ave, t, i}=s_{t,i}/g_{t,i}$. If we set a dummy weight $w_{t,i}$, whose initial value $w_{0,i}=1,\,\forall i \in \tilde{V}$, the algorithm performs in the same manner: every node works on a triplet $(s_{t,i}, g_{t,i},w_{t,i})$ and computes $\hat{x}_{ave, t, i}=\frac{(s_{t,i}/w_{t,i})}{(g_{t,i}/w_{t,i})}$.  $(s_{t,i}/w_{t,i})$ is the estimate of the average local sum on a root and  $g_{t,i}/w_{t,i}$ is the estimate of the average size of a  tree. Their relative errors are bounded in the same way as follows.

The relative error in the contributions (with respect to the diffusion effect of gossip) at node $i$ at time $t$ is $\Delta_{t,i}=\max_j |\frac{y_{t,i,j}}{\|\mathbf{y}_{t,i}\|_1}-\frac{1}{m}|=\|\frac{\mathbf{y}_{t,i}}{\|\mathbf{y}_{t,i}\|_1}-\frac{1}{m}\cdot \mathbf{1}\|_{\infty}$.
The following potential function
\[\Phi_{t}=\sum_{i,j}(y_{t,i,j}-\frac{w_{t,i}}{m})^2\]
is the sum of the variance of the contributions $y_{t,i,j}$. We name the root of the largest tree
as node $z$.


To prove Theorem~\ref{thm:gossip-ave-upper-bound}, we  need some auxiliary lemmas.

\begin{lem}[Geometric convergence of $\Phi$]\label{lem:gossip-convergence rate}
The conditional expectation $$E[\Phi_{t+1}|\Phi_{t}=\phi]=\frac{1}{2}(1-\sum_{i\in \tilde{V}} P_i^2)\phi<\frac{1}{2}\phi$$
where $P_i=(1-\delta)\frac{g_i}{n}$ is the probability that the root node $i$ is selected by any other root node, $g_i$ is the size of the tree rooted at node $i$, $\delta$ is the probability that a message fails to reach its destined root node, and $n$ is the total number of nodes in the network.
\end{lem}
\vspace*{-0.1in}
%
%
%
%

\proof
This proof is generalized from~\cite{kempe}. The difference is that the selection probability, $P_i$, is not uniform any more but depends on the tree size, $g_i$.
 $P_i$ is the probability that root $i$ is selected by any other root and $\sum_{i\in \tilde{V}}P_i^2$ is the probability that two roots select the same root.
The conditional expectation of potential at round $t+1$ is
\begin{align*}
E[&\Phi_{t+1}|\Phi_{t}=\phi]\\
&=
\frac{1}{2}\phi +\frac{1}{2}\sum_{i,j,k}\left( y_{i,j}-\frac{w_{i}}{m}\right)\left(y_{k,j}-\frac{w_{k}}{m}\right)P_i\\
&\quad +\frac{1}{2}\sum_{j,k}\sum_{k'\neq k}\left( y_{k,j}-\frac{w_{k}}{m}\right)\left(y_{k',j}-\frac{w_{k'}}{m}\right)\sum_{i\in \tilde{V}}P_i^2\\
&=
\frac{1}{2}\phi +\frac{1}{2}\sum_{i,j,k}\left( y_{i,j}-\frac{w_{i}}{m}\right)\left(y_{k,j}-\frac{w_{k}}{m}\right)P_i\\
&\quad +\frac{\sum_{i\in \tilde{V}}P_i^2}{2}\sum_{k,j,k'}\left( y_{k,j}-\frac{w_{k}}{m}\right)\left(y_{k',j}-\frac{w_{k'}}{m}\right)\\
&\quad -\frac{\sum_{i\in \tilde{V}}P_i^2}{2}\sum_{k,j}\left( y_{k,j}-\frac{w_{k}}{m}\right)^2\\
&=\frac{1}{2}(1-\sum_{i\in \tilde{V}}P_i^2)\phi \\
&\quad +\frac{1}{2}\sum_{i,j}(P_i+\sum_{i\in \tilde{V}}P_i^2)\left( y_{i,j}
 -\frac{w_{i}}{m}\right)\sum_{k}\left( y_{k,j}-\frac{w_{k}}{m}\right)\\
&=\frac{1}{2}(1-\sum_{i\in \tilde{V}}P_i^2 )\phi
<\frac{1}{2}\phi.
\end{align*}
The last equality follows from the fact that
\[\sum_{k}\left( y_{k,j}-\frac{w_{k}}{m}\right)=\sum_{k} y_{k,j}- \sum_{k}\frac{w_{k}}{m}=1-1=0.\]
\endproof
%

%
%
%
\begin{lem}\label{lem:lower-bound-weight}
There exists a $\tau=O(\log m)$ such that after $\forall t>\tau$ rounds of Gossip-ave, $w_{t,z}\geq 2^{-\tau}$ at $z$, the root of the largest tree.
\end{lem}
\vspace*{-0.1in}

\proof
 In the case that the selection probability is  uniform, it has been shown in~\cite{kempe}  that on an $m$-clique, with probability at least $1-\frac{\delta'}{2}$, after $4\log m +\log {2}{\delta'}$ rounds, a message originating from any node (through a random walk on the clique) would have  visited all nodes of the clique. When the distribution of the selection probability is not uniform, it is clear that a message originating from any node must have visited the node with the highest selection probability  after a certain number of rounds that is greater than $4\log m +\log {2}{\delta'}$ with probability at least $1-\frac{\delta'}{2}$.
\endproof

From the previous two lemmas, we derive the following theorem.

\begin{thm}[Diffusion speed of Gossip-ave]\label{thm:diffusion-speed}
With probability at least $1-\delta'$, there exists a time $T_{ave}=O(\log m +\log \frac{1}{\epsilon} +\log \frac{1}{\delta'})$, such that $\forall t \ge T_{ave}$, the contributions at  $z$, root of the largest tree, is nearly uniform, i.e.,
$\max_j|\frac{y_{t,z,j}}{\|\mby_{t,z}\|_1}-\frac{1}{m}|=\|\frac{\mathbf{y}_{t,i}}{\|\mathbf{y}_{t,i}\|_1}-\frac{1}{m}\cdot \mathbf{1}\|_{\infty} \le \epsilon$.
\end{thm}
\proof
By Lemma~\ref{lem:gossip-convergence rate}, we obtain that $E[\Phi_t]<(m-1)2^{-t}<m2^{-t}$, as $\Phi_0=(m-1)$. By Lemma~\ref{lem:lower-bound-weight}, we set $\tau=4\log m +\log \frac{2}{\delta'}$ and $\hat{\epsilon}^2=\epsilon^2\cdot \frac{\delta'}{2}\cdot 2^{-2\tau}$. Then after
$t=\log m + \log \frac{1}{\hat{\epsilon}}$ rounds of Gossip-ave,  $E[\Phi_t]\le \hat{\epsilon}$. By Markov's inequality \cite{random-alg-book}, with probability at least $1-\frac{\delta'}{2}$, the potential $\Phi_t\le \epsilon^2\cdot 2^{-2\tau}$, which guarantees that $|y_{t,i,j}-\frac{w_{t,i}}{m}|\le \epsilon \cdot 2^{-\tau}$ for all the root nodes $i$.

To have  $\max_j|\frac{y_{t,z,j}}{\|\mby_{t,z}\|_1}-\frac{1}{m}|\le \epsilon$, we need to lower bound  the weight of node $z$. From Lemma~\ref{lem:lower-bound-weight}, $w_{t,z}=\|\mby_{t,z}\|_1\geq 2^{-\tau}$ with probability at least $1-\frac{\delta'}{2}$. Note that Lemma~\ref{lem:lower-bound-weight} only applies to  $z$, the root of the largest tree. (A root node of a relatively small  tree may not be selected often enough to have such a lower bound on its weight.)
Using union bound, we obtain, with probability at least $1-\delta'$,
\(\max_j|\frac{y_{t,z,j}}{\|\mby_{t,z}\|_1}-\frac{1}{m}|\le \epsilon.\)
\endproof

Now we are ready to prove Theorem~\ref{thm:gossip-ave-upper-bound}.\\
\textbf{Proof of Theorem~\ref{thm:gossip-ave-upper-bound}}\\
\proof
From Theorem~\ref{thm:diffusion-speed}, with probability at least $1-\delta'$, it is guaranteed that after $T_{ave}=O(2\log m +\log \frac{1}{\epsilon} +\log \frac{1}{\delta'})$ rounds of Gossip-ave, at  $z$, the root of the largest tree, $\|\frac{\mathbf{y}_{t,i}}{\|\mathbf{y}_{t,i}\|_1}-\frac{1}{m}\cdot \mathbf{1}\|_{\infty}\le \frac{\epsilon}{m}$.
Let both $\epsilon= n^{-\alpha}$ and $\delta'= n^{-\alpha}, \;\alpha >0$, then $T_{ave}=O(2\log m + 2\alpha \log n)=O(\log n)$.

Using  H\"{o}lder's inequality, we obtain
\begin{align*}
&\frac{\left|\frac{s_{t,z}}{w_{t,z}}-\frac{1}{m}\sum_j x_j\right|}{\left|\frac{1}{m}\sum_j x_j\right|}
=
\frac{\left|\frac{\mby_{t,z}\cdot \mbx}{\|\mby_{t,z}\|_1} - \frac{1}{m}\cdot \mathbf{1}\cdot \mbx \right|}{\left|\frac{1}{m}\sum_j x_j\right|}
=m \cdot \frac{\left|\left(\frac{\mby_{t,z}}{\|\mby_{t,z}\|_1} - \frac{1}{m}\cdot \mathbf{1}\right)\cdot \mbx \right|}{\left|\sum_j x_j\right|}
\\
&\quad\quad \leq m \cdot \frac{\|\frac{\mby_{t,z}}{\|\mby_{t,z}\|_1} - \frac{1}{m}\cdot \mathbf{1}\|_{\infty}\cdot \|\mbx\|_1}{\left|\sum_j x_j\right|}\\
&\quad\quad \leq \epsilon \cdot \frac{\sum_j |x_j|}{\left|\sum_j x_j\right|}.
\end{align*}
When all $x_j$   have the same sign, we have
\(\frac{\left|\frac{s_{t,z}}{w_{t,z}}-\frac{1}{m}\sum_j x_j\right|}{\left|\frac{1}{m}\sum_j x_j\right|}\leq \epsilon.\)
Further, we need to bound the relative error of  $Ave$.
W. l. o. g., let the \emph{true} average of the sum of values in a  tree be positive, i.e., $s_{ave}=\frac{1}{m}\sum_j x_j >0$ and, by definition, the \emph{true} average of the size of a tree is also positive, i.e., $g_{ave}=\frac{1}{m}\sum_j g_j=\frac{n}{m} >0$.
Therefore, the global average $Ave$ is  $x_{ave}=\frac{s_{ave}}{g_{ave}}$. Since   $|\frac{s_{t,z}}{w_{t,z}}-s_{ave}|\leq \epsilon s_{ave}$ and $|\frac{g_{t,z}}{w_{t,z}}-g_{ave}|\leq \epsilon g_{ave}$ we obtain
\begin{align*}
\hat{x}_{ave, t z}&=\frac{s_{t,z}}{g_{t,z}}=\frac{\left(\frac{s_{t,z}}{w_{t,z}}\right)}{\left(\frac{g_{t,z}}{w_{t,z}}\right)}
\in \left[\frac{1-\epsilon}{1+\epsilon}\frac{s_{ave}}{g_{ave}} ,\;\;\, \frac{1+\epsilon}{1-\epsilon}\frac{s_{ave}}{g_{ave}}\right].
\end{align*}

Set $\epsilon'=c\epsilon$, where $c=\frac{2}{(1-\epsilon)}>2$ is bounded when $\epsilon <1$. (For example, if $\epsilon \leq 10^{-2}$, then $c=2.\bar{0}\bar{2}$ and
$\epsilon'=\frac{200}{99}\epsilon$.) We set $\epsilon=n^{-\alpha}$, and then $\epsilon'=\frac{2}{n^{\alpha}-1} \approx 2\epsilon$.
Thus, with probability at least $1-\frac{1}{n^{\alpha}}$, the relative error at $z$ is
\[\frac{|\hat{x}_{ave, t, z}- x_{ave}|}{|x_{ave}|}\leq \epsilon',\]
 after at most  $O(\log m + 2\alpha\log n)=O(\log n)$ rounds of Gossip-ave algorithm.
%

The above assumption that all $x_j$ have the  same sign is just for complexity analysis but not for the execution of the gossip-ave algorithm. The gossip-ave algorithm works well without any assumption 
on the values of  roots. In the following, we further relax this assumption and show that the upper bound on the running time is also valid when  $x_j$ are not all of the same sign.

Let $\gamma=\|\mbx\|_1\neq 0$ and $\mbx'=\mbx+2\gamma \cdot \mathbf{1} > \mathbf{0}$, i.e., all $x'_j >0$ have the same sign. It is obvious that the average aggregate of the $\mbx'$ is a simple offset of the average aggregate of the $\mbx$, i.e., $x'_{ave}=x_{ave}+2\gamma$. Proceeding through the same data exchanging scenario in each round of the gossip-ave algorithm on $\mbx$ and $\mbx'$, after $t$ rounds, at root node $z$, we have the relationship between the two corresponding estimates of the  average aggregates on $\mbx'$ and $\mbx$:  $\hat{x}'_{ave,\,t,\,z}=\hat{x}_{ave,\,t,\,z}+2\gamma$.
The desired related error is $\frac{|\hat{x}_{ave,\,t,\,z}-x_{ave}|}{|x_{ave}|}\le \epsilon'$. 
Let $\gamma=O(n^{\alpha})$ and a stricter threshold $\tilde{\epsilon}=\epsilon' \frac{|x_{ave}|}{|x_{ave}+2\gamma|}<\epsilon'$. As all $x'_j$ are of the same sign, whp at least $1-\frac{1}{\delta'}$, after $t=O(\log n + \log \frac{1}{\tilde{\epsilon}}+ \log \frac{1}{\delta'})=O(\log n)$ rounds, 
 \[\frac{|\hat{x}'_{ave,\,t,\,z}-x'_{ave}|}{|x'_{ave}|}=\frac{|(\hat{x}_{ave,\,t,\,z}+2\gamma)-(x_{ave}+2\gamma)|}{|x_{ave}+2\gamma|}\le \tilde{\epsilon}= \epsilon' \frac{|x_{ave}|}{|x_{ave}+2\gamma|}.\]
 From the above equation, we conclude that  \[\frac{|\hat{x}_{ave,\,t,\,z}-x_{ave}|}{|x_{ave}|}\le \epsilon'.\]     
That is to say, running the gossip-ave algorithm on an arbitrary vector $\mbx$, whp at least $1-\frac{1}{\delta'}$, after $t=O(\log n + \log \frac{1}{\tilde{\epsilon}}+ \log \frac{1}{\delta'})=O(\log n)$ rounds, the relative error of the estimate of the average aggregate  is less than  $\epsilon'=\frac{2}{n^{\alpha}-1}=O(n^{-\alpha})$.

\endproof
Evaluating the performance of the gossip-ave algorithm using the criterion of relative error causes a problem when $x_{ave}=0$ whereas the gossip-ave algorithm works well when $x_{ave}=0$. In this case, using  absolute error criterion, i.e. $|\hat{x}_{ave,\,t,\,z}-x_{ave}|=
|\hat{x}_{ave,\,t,\,z}|\le \epsilon'$ is more suitable. Here, we would show that the upper bound of running time of Theorem~\ref{thm:gossip-ave-upper-bound} is also valid for the case that $x_{ave}=0$ and the performance is assessed under the absolute error criterion $|\hat{x}_{ave,\,t,\,z}|\le \epsilon'$. By the similar technique as in the above proof, choose an offset constant $\gamma= O(n^{\alpha}) >1$ such that $\mbx'=\mbx+\gamma \cdot \mathbf{1} > \mathbf{0}$, i.e., all $x'_j >0$ are with the same sign. Also, let $\tilde{\epsilon}=\frac{\epsilon'}{|x'_{ave}|}=\frac{\epsilon'}{\gamma}<\epsilon'$. Proceeding through the same data exchanging scenario in each round of the gossip-ave algorithm on $\mbx$ and $\mbx'$, whp at least $1-\frac{1}{\delta'}$, after $t=O(\log n + \log \frac{1}{\tilde{\epsilon}}+ \log \frac{1}{\delta'})=O(\log n)$ rounds,       
\[\frac{|\hat{x}'_{ave,\,t,\,z}-x'_{ave}|}{|x'_{ave}|}=\frac{|(\hat{x}_{ave,\,t,\,z}+\gamma)-\gamma|}{\gamma}\le \tilde{\epsilon}= \frac{\epsilon'}{\gamma}.\]
 From the above equation, we have that  \(|\hat{x}_{ave,\,t,\,z}|\le \epsilon'.\)
 This concludes the mapping relationship between the relative error criterion and the absolute error criterion.

\noindent {\bf Complexity of Gossip-ave} \\
Gossip-ave algorithm needs $O(\log m + \log \frac{1}{\epsilon})=O(\log n)$ rounds and $m\cdot O(\log n)=O(n)$ messages for the root of the largest tree to have the global average aggregate, $Ave$, within a relative error of at most $\frac{2}{n^{\alpha}-1},\; \alpha >0$.
\vspace*{-0.15in}
\subsection{DRR-gossip Algorithms}\label{sec:whole algorithm}
\vspace*{-0.1in}
Putting together our results from the previous subsections, we present 
 Algorithm~\ref{alg:DRR-gossip-max},  DRR-gossip-max algorithm, and  Algorithm~\ref{alg:DRR-gossip-ave},  DRR-gossip-ave algorithm, for computing $Max$ and  $Ave$, respectively.
 To conclude from  previous sections, the time complexity of  DRR-gossip  is $O(\log n)$ since all  phases need $O(\log n)$ rounds. The message complexity is dominated by  DRR algorithm in  phase I which  needs $O(n\log\log n )$ messages.

The DRR-gossip-ave algorithm is more involved than the DRR-gossip-max
algorithm.
Unlike the Gossip-max algorithm which ensures that all the roots will have
$Max$ whp, the Gossip-ave algorithm only guarantees that the root of the
largest tree in terms of tree size will have the $Ave$ whp. To ensure that all
the roots have $Ave$ whp, after the Gossip-ave algorithm, the root of the
largest tree has to spread out its estimate, the $Ave$, by using the Data-
spread algorithm where the root of the largest tree sets its estimate, the
$Ave$, computed by the Gossip-ave algorithm, as the data to be spread out.
Therefore, every root needs to know in advance whether it is the root of the
largest tree. To achieve this, the Gossip-max algorithm is executed beforehand
on tree sizes which are obtained from the Convergecast-sum algorithm.
(Note that the Gossip-max procedure in the DRR-gossip-max algorithm is
executed on the local maximums computed by the Convergecast-max algorithm.)
Every root could compare the maximum tree size
obtained from the Gossip-max algorithm with the size of its own tree to recognize
whether it is the root of the largest tree.
(Note that the Gossip-max algorithm and the Gossip-ave algorithm can not be
executed simultaneously, since the Gossip-ave algorithm does not have the
sampling procedure as in the Gossip-max algorithm.)
    Finally, every root then broadcasts the  $Ave$ obtained from the
Data-spread algorithm to all its tree members.

%
%
%
\begin{algorithm}
\label{alg:DRR-gossip-max}
\caption{DRR-gossip-max}
Run $DRR(G)$ to obtain the forest $\mathbb{F}$.\;
Run Convergecast-max($\mathbb{F}$,$\mathbf{v}$).\;
Run Gossip-max($G,\,\mathbb{F},\,\tilde{V},\,\mathbf{cov}_{max}$).\;
Every root node broadcasts the $Max$ to all nodes in its  tree.\;
\end{algorithm}
%
%
\begin{algorithm}
\label{alg:DRR-gossip-ave}
\caption{DRR-gossip-ave}
Run $DRR(G)$ algorithm to obtain the forest $\mathbb{F}$.\;
Run Convergecast-sum$(\mathbb{F}$, $\mathbf{v})$ algorithm.\;
Run Gossip-max$(G,\,\mathbb{F},\,\tilde{V},\,\mathbf{cov}_{sum}(*,2))$ algorithm on the sizes of trees to find the root of the largest tree. At the end of this phase, a root $z$ will know that it is the one with the largest tree size.\;
Run Gossip-ave$(G,\,\mathbb{F},\,\tilde{V},\,\mathbf{cov}_{sum})$ algorithm.\;
Run Data-spread$(G,\,\mathbb{F},\,\tilde{V},\,Ave)$ algorithm---the root of the largest  tree uses its average estimate, i.e., $Ave$, as the value to spread.\;
Every root broadcasts its value to all the nodes in its  tree.
\end{algorithm}

\vspace*{-0.1in}
\subsection{The complexity of DRR-gossip algorithms}
\vspace*{-0.05in}
To conclude from the previous sections, the time complexity of the DRR-gossip algorithms is $O(\log n)$ since all the phases need $O(\log n)$ rounds. The message complexity is dominated by the DRR algorithm in the phase I which  needs $O(n\log\log n )$ messages.
Thus, our DRR-gossip algorithms achieve the same time complexity as uniform gossip of~\cite{kempe} but reduce the message complexity to $O(n\log\log n)$. Although the efficient gossip of~\cite{efficient-gossip} can have the same message complexity, it will need $O(\log \log\log n)$ time.


\vspace*{-0.15in}
\section{Application to  Sparse Networks --- Local-DRR Algorithm}
\label{sec:p2p}
\vspace*{-0.1in}
%
%
In  sparse networks, a small
number of neighbors  makes it feasible for each node to send  messages to
all of its neighbors simultaneously in one round. In fact, this is a standard assumption in the traditional message passing distributed computing model \cite{peleg} (here it is assumed  messages sent to different
neighbors in one round can all be different). We show how DRR-gossip can be used to
improve  gossip-based aggregate computation in such networks.

 We assume that, in a round of time, a node of an arbitrary undirected graph can communicate
  directly only with its immediate neighbors (i.e., nodes that are connected directly by an edge). (Note that, in previous sections, any two nodes can communicate with each other in a round under a complete graph model.) Thus, on such a communication model, we have a variant of the DRR algorithm, called the {\em Local-DRR algorithm}, where a node only exchange rank information with its immediate neighbors.
  Each node chooses a random rank in $[0,1]$ as before.
  Then each node connects  to its  highest ranked neighbor (i.e., the neighbor which has
   the highest rank among all its neighbors).   A node that has the highest rank among
  all its neighbors will become a root. Since every node, except root nodes, connects
  to a node with higher rank, there is no cycle in the graph.   Thus this process
  results in a collection of  disjoint  trees. As shown  in Theorem~\ref{th:height} below, the key property
  is that the {\em height} of each tree produced by the Local-DRR algorithm on an {\em arbitrary} graph is
   bounded by $O(\log n)$ whp. This enables us to bound the time
   complexity of the Phase  II of the DRR-gossip algorithm, i.e., Convergcast and Broadcast, on an arbitrary graph by $O(\log n)$ whp.

\begin{thm}
\label{th:height}  On an arbitrary undirected graph, all the trees produced by the Local-DRR algorithm have
a height of at most $O(\log n)$ whp.
\end{thm}
%
\vspace*{-0.1in}
\proof Fix any  node $u_0$. We first show that
the path from $u_0$ to a root  is at most $O(\log n)$ whp.
 Let $u_1, u_2, \dots$ be the successive ancestors of $u_0$, i.e.,
$u_1$ is the parent of $u_0$ (i.e., $u_0$ connects to $u_1$), $u_2$ is the parent of $u_1$ and so
on. (Note $u_1, u_2, \dots$ are all null if $u_0$ itself is the
root).  Define the complement value to the rank of $u_i$ as $C_i
:= 1 - rank(u_i)$, $i \geq 0$.  The main thrust of the proof is to
show that the sequence $C_i$, $i \geq 0$ decreases geometrically
whp. We adapt  a technique used in \cite{srin}.

For $t \geq 0$, let $I_t$ be the indicator random variable for the
event that a root has not been reached after $t$ jumps, i.e.,
$u_0, u_1, \dots, u_t$ are not roots.
We need the following  Lemma.

\begin{lemma}
\label{le:treeheight} For any $t \geq 1$ and any $z \in [0,1]$,
$E[C_{t+1}I_t| C_tI_{t-1} = z] \leq z/2$.
\end{lemma}
\vspace*{-0.1in}
\proof We can assume that $z \neq 0$; since $C_{t+1} \leq C_t$ and
$I_t \leq I_{t-1}$,  the lemma holds trivially if $z = 0$.
Therefore, we have $I_{t-1} = 1$ and $C_t = z > 0$. We focus on
the node $u_t$.   Denote the set of neighbors of node $u_t$ by  $U$; the size of $U$ is  at most $n -
1$.  Let $Y$ be the random variable denoting the number of
``unexplored" nodes  in set
 $U$, i.e., those that do not belong to the set $\{u_0, u_1, \dots, u_{t-1}\}$. If $Y=0$, then $u_t$ is a root and
 hence $C_{t+1}I_t = 0$. We will prove that for all $d \geq 1$,
 \begin{equation}
 \label{eq:cond}
 E[C_{t+1}I_t| ((C_tI_{t-1} = z) \wedge (Y = d))] \leq z/2.
 \end{equation}

 Showing the above is enough to prove the lemma, because if the lemma holds conditional on
 all positive values of $d$, it also holds unconditionally. For convenience, we denote the l.h.s. of (\ref{eq:cond}) as
 $\Phi$.

 Fix some $d \geq 1$. In all arguments below, we condition on the event ``$(C_tI_{t-1} = z) \wedge (Y = d)"$.
 Let  $v_1, v_2, \dots, v_d$ denote the $d$ unexplored nodes in $U$. If $rank(v_i) < rank(u_t)$ for all $i$ ($1 \leq i \leq d$), then
 $u_t$ is a root and hence $C_{t+1}I_t = 0$. Therefore, conditioning on the value $y = \min_{i} C_i = min_{i} (1 - rank(v_i))  \leq z$, and considering the $d$ possible values of $i$ that achieve this minimum, we get,
 $$\Phi = d \int_{0}^z y (1-y)^{d-1} dy.$$

 Evaluating the above yields
\begin{equation*}
\label{eq:int} \Phi = \frac{1 - (1 - z)^d (1 + zd)}{(d+1)}.
\end{equation*}

We can show that the r.h.s of the above is at most $z/2$ by a straightforward induction on $d$.
\endproof

Using Lemma \ref{le:treeheight}, we now prove Theorem
\ref{th:height}.

We have $E[C_1I_0] \leq E[C_1] \leq 1$.  Hence by Lemma
\ref{le:treeheight} and an induction on $t$ yields  that
$E[C_tI_{t-1}] \leq 2^{-t}$. In particular, letting $T = 3\log n$,
where $c$ is some suitable constant, we get
$E[C_TI_{T-1}] \leq n^{-3}.$

Now, suppose $u_T = u$  and that $C_TI_{T-1} = z$.  The degree of node $u$
 is   at most $n$; for each of these nodes $v$,
$\Pr(rank(v) > rank(u))=\Pr(1-rank(v) < 1-rank(u))=\Pr(1-rank(v) <
z)=z$. Thus the probability that $u$ is not a root is at most
$n z$; more formally,
$\forall z, \Pr(I_T = 1 |C_T I_{T-1} = z) \leq  n z.$
So,
$$\Pr(I_T = 1) \leq \log n  E[C_TI_{T-1}] \leq n /n^3 = 1/n^2.$$

Hence, whp, the number of hops from any fixed note to the root is
$O(\log n)$.  By union bound, the statement holds for all nodes
whp.
\endproof

Similar to Theorem~\ref{le:numtrees}, we can bound the number of trees produced
by the Local-DRR algorithm on an arbitrary graph.

\begin{thm}
\label{th:numtrees}
Let $G$ be an arbitrary connected undirected graph having $n$ nodes.
Let $d_i = O(n/\log n)$ be the degree of node $i$, $1 \leq i \leq n$.
The  number of  trees produced by the Local-DRR algorithm is $O(\sum_{i=1}^n \frac{1}{d_i +1})$ whp.
Hence, if $d_i = d$, $\forall i$, then the number of trees  is $O(n/d)$ whp.
\end{thm}
\vspace*{-0.1in}
\proof
Let the indicator random variable $X_i$  take the value of 1 if node $i$ is
a root and 0 otherwise.
Let $X = \sum_{i=1}^n X_i$ be the total number of roots.
$\Pr(X_i = 1) = 1/(d_i + 1)$ since, this is the probability its value is the highest among all of its
$d_i$ neighbors. Hence, by linearity of expectation, the expected number of
roots (hence, trees) is
$E[X] = \sum_{i=1}^n E[X_i] =  \sum_{i=1}^n \frac{1}{d_i +1}.$ To show concentration, we cannot directly use a standard Chernoff bound since $X_i$s are not independent (connections are not independently chosen, but fixed by the underlying graph). However, one can use the following variant of the
Chernoff bound from \cite{Srinivasan} (cf. Lemma \ref{lem_fkg}),
which works in the case of
dependent  indicator random variables that are  correlated as
defined below. For  random variables, $X_1, \dots, X_i, \dots, X_n$ and for any $S_{i-1} \subseteq
\{1,\ldots,i-1\}, \Pr(X_i = 1 | \bigwedge_{j \in S_{i-1}} X_j=1)
\leq \Pr(X_i = 1)$. This is because if a node's neighbor is a root, then the probability that the node
itself is a root is 0. Also, the assumption of $d_i=O(n/\log n)$ ensures that $E[X]$ is $\Omega(\log n)$, so the Chernoff bound yields a high probability  on the concentration of $X$ to its mean $E[X]$.
\endproof


We make two assumptions regarding the network communication model:
(1) as mentioned earlier, a node can  send a message simultaneously to all its neighbors (i.e., nodes that  are connected directly by an edge) in the same
round; (2) there is a routing protocol which allows any node to
 communicate with a {\em random} node in the network in $O(T)$
rounds and using $O(M)$ messages whp. Assumption (1) is standard in distributed computing literature\cite{DRG,peleg}. As for Assumption (2), there are well-known techniques for
sampling a random node in a network, e.g., using random walks (e.g., \cite{zhong})  or
using special properties of the underlying topology, e.g., as in
P2P topologies such as Chord \cite{saia}.
Under the above assumptions, we obtain the performance of DRR-gossip using the Local-DRR algorithm on sparse graphs in the following Theorem.

\begin{thm}\label{thm:sparse-complexity}
On a $d$-regular graph $G(V,E)$, where $|V|=n$ and $d = O(n/\log n)$, the time complexity of the DRR-gossip algorithms is $O(\log n + T\log \frac{n}{d})$ whp by using the Local-DRR algorithm and a routing protocol running in $O(T)$ rounds and $O(M)$ messages (whp) between a gossip pair; the corresponding message complexity
is $O(|E| + \frac{n}{d}M \log \frac{n}{d})$ whp.
\end{thm}
\vspace*{-0.1in}
\proof
Phase I (Local-DRR) takes $O(1)$ time, since each node can find its largest ranked neighbor in constant time (Assumption 1) and needs $O(|E|)$ messages in total (since at most two messages travel through an edge). Phase II (convergecast and broadcast) takes $O(\log n)$ time (by Theorem \ref{le:treeheight} and Assumption 1) and $O(n)$ messages. Phase III (uniform gossip) takes $O(T\log \frac{n}{d})$  time (Assumption 2) and needs $O(\frac{n}{d}M \log \frac{n}{d})$ messages (Assumption 2 and Theorem \ref{th:numtrees}).
\endproof

We can apply the above theorem to Chord  \cite{chord}. Each node in Chord has a degree $d = O(\log n)$.  Chord admits  an efficient  (non-trivial) protocol (cf. \cite{saia}) which satisfies Assumption (2) with $T = O(\log n)$ and $M = O(\log n)$ (both in expectation, which is sufficient here). Hence the above theorem shows that DRR-gossip takes $O(\log^2 n)$ time  and $O(n\log n)$ messages whp. In contrast,  the straightforward uniform gossip \cite{kempe} gives $O(T\log n)=O(\log^2 n)$ rounds and $O(M\cdot n \log n)=O(n \log^2 n)$ messages whp. 

\vspace*{-0.15in}
\section{Lower Bound for Address-Oblivious Algorithms}
\label{sec:lbound-add-oblivious}
\vspace*{-0.1in}
We conclude  by  showing a non-trivial lower bound result on gossip-based aggregate computation: any address-oblivious algorithm for computing aggregates requires $\Omega(n \log n)$ messages, {\em regardless of the number of rounds or the size of the (individual) messages}.
We assume the random phone call model: i.e., communication partners are chosen randomly (without depending on their addresses).  The following theorem gives a lower bound for computing the Max aggregate. The  argument can be adapted for other aggregates as well.

\vspace{-0.15in}
\begin{thm}
Any address-oblivious algorithm that computes  the Maximum value, $Max$, in a  $n$-node network  needs $\Omega(n \log n)$ messages whp (regardless of the number of rounds).
\end{thm}
\vspace{-0.15in}

\noindent \proof
We lower bound  the number of messages   exchanged between nodes before a large fraction of the nodes correctly
knows the (correct) maximum value. Suppose nodes can send messages that are arbitrary long. (The bound will hold regardless of this assumption.)  Without loss of generality, we will assume that a node can send a  list of all node addresses and the corresponding node values learned so far (without any aggregation). For any  node $i$ to have correct knowledge of the maximum, it should
somehow know  the values at all other nodes.  (Otherwise,  an adversary ---who knows the random choices made by the algorithm --- can always make sure that the maximum is at a node which is not known by $i$.) There are two ways that $i$ can learn about  another node $j$'s value: (1) direct way:  $i$ gets to
know $j$'s value by communicating with $j$ directly (at the beginning, each node knows only about its own value); and (2)  indirect way: $i$ gets to know $j$'s value by communicating with a node $w \neq j$ which has a knowledge of $j$'s value. Note that $w$ itself  may have learned
 about $j$'s value either directly or indirectly.

Let $v_i$ be the (initial) value associated with node $i$, $1\leq  i \leq n$.  We will assume that all values are {\em distinct}.
By the adversary argument,  the requirement is that at the end  of any algorithm, on the average, at least half  of the nodes should know  (in the above direct or indirect way)  all  of the $v_i$, $1\leq  i \leq n$.
 Otherwise, the adversary can make that value that is not known to more than half of the nodes, the maximum.
We want to show that the number of messages needed to satisfy the above requirement is at least  $c n \log n$, for some (small) constant $c > 0$. In fact, we show something stronger:  at least  $c n \log n$ (for some small $c > 0$) messages are needed if we  require even
 $n^{\Omega(1)}$ values  to be known to at least $\Omega(n)$  nodes.

We define a
  stage (consisting of one or more rounds) as follows.  Stage 1 starts with  round 1. If stage  $t$ ends in round $j$, then  stage $t + 1$ starts in round $j+ 1$. Thus, it remains
to describe when a stage ends. We distinguish sparse and dense stages. A sparse stage contains at most $\epsilon n$
messages (for a suitably chosen small constant $\epsilon > 0$, fixed later in the proof). The length of these stages is maximized,  i.e., a sparse stage ends in a  round $j$ if adding round $j+1$ to the stage would result in more than
$\epsilon n$ messages. A dense stage consists of only one round containing more than $\epsilon n$ messages. Observe that the number of messages
during the stages 0 to $j$ is at least $(j-1)\epsilon n/2$ because any pair of consecutive stages contains at least $\epsilon n$ messages by construction.

Let $S_i(t)$ be the set of {\em nodes} that  know $v_i$ at the beginning of stage $t$.
At the beginning of stage 1, $|S_i(1)| = 1$, for all $1 \leq i \leq n$.

At the beginning of stage  $t$, we call a value  as {\em  typical} if  it is known by at most $6^t \log n$ nodes (i.e.,
$|S_i(t)| \leq 6^t \log n $) {\em and} it  was typical at the beginning of all stages prior to $t$. All values are typical
at the beginning of stage 1. 
Let $k_t$ denote the number of typical values  at the beginning of stage $t$. 

 The proof of the Theorem follows from the following claim. (Constants specified will be fixed in the proof; we don't try to optimize  these values).

{\bf Claim:} At the beginning of stage $t$, at least  $(1/6)^t n$  values are typical  w.h.p., for all $t \leq \delta \log n$, for a fixed positive constant $\delta$. 


The above claim will imply the theorem since
at the end of stage $t = \delta \log n$,  $|S_i(t)| \leq o(n)$  for at least $n^{\Omega(1)}$ values, i.e., at least
 $n^{\Omega(1)}$ values are not yet known to $1- o(1)$ fraction of the nodes after stage $t =  \delta \log n$.
 Hence the number of messages needed is at least $\Omega(n \log n)$.


 We prove the above claim by induction: We show that if the claim holds at the beginning of a stage then it
 hold at the end of the stage. We show this regardless whether  the stage  is   dense or sparse, and thus we have two cases.

\noindent {\em Case 1:}  The stage is dense. A dense stage consists of only one round with at least $\epsilon n$ messages.
Fix a typical value $v_i$.
Let $U_i(t) = V - S_i(t)$, i.e., the set of nodes  that do not know $v_i$ at the beginning of stage $t$.
For $1 \leq k(i) \leq |U_i(t)|$, let $x_{k(i)}$ denote the indicator random variable that denotes whether the $k(i)$th of these nodes gets to know the value $v_i$ in this stage. Let $X_i(t) =  \sum_{k(i) = 1}^{|U_i(t)|}x_{k(i)} $.
Let $u$ be a node that does not know $v_i$.
$u$ can get to know $v_i$ either by calling a node that knows the value or being called by a node
that knows the value.
 The probability it gets to know $v_i$ by calling is at most $6^t \log n/n$ and the probability that it gets called by
 a node knowing the value is at most $6^t \log n/n$ (this quantity is $o(1)$, since $ t \leq  \delta \log n$ and $\delta$ is sufficiently small). Hence the total probability that it gets to know $v_i$ is at most $2\cdot 6^t \log n/n$.
Thus, the expected number of nodes that get to know $v_i$ in this stage  is $ E[X_i(t)] = \sum_{k(i) = 1}^{|U_i(t)|} \Pr\{x_{k(i)} = 1\} \leq 2\cdot 6^t \log n$.
The variables $x_{k(i)}$ are not independent, but are negatively  correlated in the sense of  Lemma \ref{lem_fkg} and using the Chernoff bound of this Lemma  we have:

$\Pr(X_i(t) > 5\cdot6^t \log n) = \Pr(X_i(t) > (1 + 3/2 )\cdot 2\cdot 6^t \log n)  \leq  1/n^2.$

By union bound, w.h.p., at most $5\cdot6^t$ new nodes get to know each typical value.
Thus w.h.p. the total number of nodes knowing a typical value (for every such value)  in this stage is at most $6^t \log n + 5 \cdot6^t \log n
= 6^{t+1}\log n$, thus satisfying the induction hypothesis. It also follows that a typical value at the beginning of a dense phase remains typical at the end of the phase, i.e., $k_{t+1} = k_t$ w.h.p.

\noindent {\em Case 2:} The stage is sparse. By definition, there are at most $\epsilon n $ messages
in a sparse stage.  Each of these messages can be a push or a pull. A sparse stage may consist of multiple rounds.

Fix a typical
value $v_i$.  W.h.p, there are at most $6^t\log n$ nodes that know a typical value at the beginning of this stage.
Using pull messages, since the origin  is chosen uniformly at random, the probability that one of these nodes is contacted
is at most $1/n (\epsilon n) = \epsilon$.  Hence the expected number of messages sent by nodes knowing this typical value
is at most $\epsilon 6^t \log n$.  Thus the expected number of new nodes that get to know this typical value is
at most $\epsilon  6^t \log n$.  The high probability bound can be shown as earlier.

We next consider the effect of  push messages. We focus on values that are typical at the beginning of this stage. We show that
  high probability at least some constant fraction of the typical values remain typical at the end of this phase. As defined earlier, let $k_t$ be the number of such typical values. In this stage, at most $\epsilon n$ nodes are involved
  in pushing --- let this set be $Q$. Consider a random typical value $x$.
Since a typical value is known by at most $6^t \log n$  nodes and
destinations are uniformly randomly chosen, the probability that $x$ is known to
a node in $Q$ is $O(\frac{6^t \log n}{n})$. Hence the expected number of times
that $x$ will be pushed by set $Q$ is at most $O(\epsilon 6^t \log n)$. Now,
the number of times $x$   has to be pushed is
at least $(6 - \epsilon)\cdot 6^t \log n$   to  exceed  the required expansion for this  value whp  (as argued in the above para, pulling only results in at most $\epsilon 6^t \log n$ messages having   being sent out w.h.p).
By Markov's inequality, the probability that $x$ is pushed more than $(6 - \epsilon)\cdot 6^t \log n$ times by nodes in set $Q$ is at most  $\frac{\epsilon}{6-\epsilon}$.  Hence the expected number of  typical values  that can expand is at most
$\frac{\epsilon}{6-\epsilon} k_t$. Thus, in expectation,  at least  
$1 - \frac{\epsilon}{6 -\epsilon}$  fraction of the typical values remain typical. High probability bound can be shown similar to case 1. We want $1 - \frac{\epsilon}{6 -\epsilon} > 1/6$, for the induction hypothesis to hold; this can be satisfied by choosing
$\epsilon$  small enough. 
\endproof

\vspace*{-0.15in}
\section{Concluding Remarks}
\label{sec:conclusion}
\vspace*{-0.1in}
We presented an almost-optimal gossip-based protocol for computing aggregates that takes $O(n \log \log n)$ messages and $O(\log n)$ rounds.   We also showed how our protocol can be applied to improve performance in  networks with a fixed underlying topology. The main technical ingredient of our approach is a simple distributed randomized procedure
called DRR to partition a network into trees of small size. 
The improved bounds come at the cost of sacrificing  address-obliviousness. However, as we show in our lower bound,  this is necessary if we need to break the 
  the $\Omega(n \log n)$ message barrier.  An interesting open question is to establish whether $\Omega(n \log \log n)$ messages is a lower bound for gossip-based aggregate computation in the non-address oblivious  model. Another interesting direction is to see whether the DRR technique
  can be used to obtain improved bounds for other distributed computing problems.


\begin{thebibliography}{99}

\bibitem{S-Boyd-gossip}
Stephen Boyd, Arpita Ghosh, Balaji Prabhakar, and Devavrat Shah,
  \emph{Randomized gossip algorithms}, IEEE Trans. on Information Theory
  \textbf{52} (2006), no.~6, 2508--2530.

\bibitem{DRG}
Jen-Yeu Chen, Gopal Pandurangan, and Dongyan Xu, \emph{Robust aggregate
  computation in wireless sensor network: distributed randomized algorithms and
  analysis}, IEEE Trans. on Paral. and Dist. Sys. (TPDS) \textbf{17} (Sep.
  2006), no.~9, 987--1000.

\bibitem{demers}
Alan Demers, Dan Greene, Carl Hauser, Wes Irish, John Larson, Scott Shenker,
  Howard Sturgis, Dan Swinehart, and Doug Terry, \emph{Epidemic algorithms for
  replicated database maintenance}, PODC, 1987, pp.~1--12.

\bibitem{geographic-gossip}
Alexandros~G. Dimakis, Anand~D. Sarwate, and Martin~J. Wainwright,
  \emph{Geographic gossip: efficient aggregation for sensor networks}, IPSN,
  2006, pp.~69--76.

\bibitem{sparse-aggregation}
Jie Gao, Leonidas Guibas, Nikola Milosavljevic, and John Hershberger,
  \emph{Sparse data aggregation in sensor networks}, IPSN, 2007, pp.~430--439.

\bibitem{Jelasity-tocs}
M\'{a}rk Jelasity, Alberto Montresor, and Ozalp Babaoglu, \emph{Gossip-based
  aggregation in large dynamic networks}, ACM Trans. Comput. Syst. \textbf{23}
  (2005), no.~3, 219--252.

\bibitem{Karp-rumor}
Richard~M. Karp, Christian Schindelhauer, Scott Shenker, and Berthold
  V\"{o}cking, \emph{Randomized rumor spreading}, FOCS, 2000, pp.~565--574.

\bibitem{efficient-gossip}
Srinivas Kashyap, Supratim Deb, K.~V.~M. Naidu, Rajeev Rastogi, and Anand
  Srinivasan, \emph{Efficient gossip-based aggregate computation}, PODS, 2006,
  pp.~308--317.

\bibitem{kempe}
David Kempe, Alin Dobra, and Johannes Gehrke, \emph{Gossip-based computation of
  aggregate information}, FOCS, 2003, pp.~482--491.

\bibitem{saia}
Valerie King, Scott Lewis, Jared Saia, and Maxwell Young, \emph{Choosing a
  random peer in chord}, Algorithmica \textbf{49} (2007), no.~2, 147--169.

\bibitem{impact-aggregate}
Bhaskar Krishnamachari, Deborah Estrin, and Stephen~B. Wicker, \emph{The impact
  of data aggregation in wireless sensor networks}, DEBS, 2002, pp.~575--578.

\bibitem{smart-gossip}
Pradeep Kyasanur, Romit~Roy Choudhury, and Indranil Gupta, \emph{Smart gossip:
  An adaptive gossip-based broadcasting service for sensor networks}, MASS,
  2006, pp.~91--100.

\bibitem{aggregation-service}
Samuel Madden, Michael~J. Franklin, Joseph~M. Hellerstein, and Wei Hong,
  \emph{Tag: a tiny aggregation service for ad-hoc sensor networks}, SIGOPS
  Oper. Syst. Rev. \textbf{36} (2002), no.~SI, 131--146.

\bibitem{random-alg-book}
Michael Mitzenmacher and Eli Upfal, \emph{Probability and computing}, Cambridge
  University Press, 2005.

\bibitem{srin}
Ruggero Morselli, Bobby Bhattacharjee, Michael~A. Marsh, and Aravind
  Srinivasan, \emph{Efficient lookup on unstructured topologies}, PODC, 2005,
  pp.~77--86.

\bibitem{separable-gossip}
Damon Mosk-Aoyama and Devavrat Shah, \emph{Computing separable functions via
  gossip}, PODC, 2006, pp.~113--122.

\bibitem{sensys04-2}
Suman Nath, Phillip~B. Gibbons, Srinivasan Seshan, and Zachary~R. Anderson,
  \emph{Synopsis diffusion for robust aggregation in sensor networks}, SenSys,
  2004, pp.~250--262.

\bibitem{Srinivasan}
A.~Panconesi and A.~Srinivasan, \emph{Randomized distributed edge coloring via
  an extension of the {C}hernoff-{H}oeffding bounds}, SIAM Journal on Computing
  \textbf{26} (1997), 350--368.

\bibitem{peleg}
David Peleg, \emph{Distributed computing: A locality-sensitive approach}, SIAM,
  2000.

\bibitem{Gossip-Roberto-podc08}
Roberto~Di Pietro and Pietro Michiardi, \emph{Brief announcement: Gossip-based
  aggregate computation: computing faster with non address-oblivious schemes},
  PODC, 2008, page 442, Extended version at
  \url{http://www.eurecom.fr/~michiard/downloads/podc08_a_ext.pdf}.

\bibitem{pittel}
Boris Pittel, \emph{On spreading a rumor}, SIAM J. Appl. Math. \textbf{47}
  (1987), no.~1, 213--223.

\bibitem{pastry}
Antony I.~T. Rowstron and Peter Druschel, \emph{Pastry: Scalable, decentralized
  object location, and routing for large-scale peer-to-peer systems},
  Middleware, 2001, pp.~329--350.

\bibitem{hierarchical-gossip}
Rik Sarkar, Xianjin Zhu, and Jie Gao, \emph{Hierarchical spatial gossip for
  multi-resolution representation in sensor netowrk}, IPSN, 2007, pp.~420--429.

\bibitem{sensys04-1}
Nisheeth Shrivastava, Chiranjeeb Buragohain, Divyakant Agrawal, and Subhash
  Suri, \emph{Medians and beyond: new aggregation techniques for sensor
  networks}, SenSys, 2004, pp.~239--249.

\bibitem{chord}
Ion Stoica, Robert Morris, David Karger, M.~Frans Kaashoek, and Hari
  Balakrishnan, \emph{Chord: A scalable peer-to-peer lookup service for
  internet applications}, SIGCOMM, 2001, pp.~149--160.

\bibitem{zhong}
Ming Zhong and Kai Shen, \emph{Random walk based node sampling in
  self-organizing networks}, SIGOPS Oper. Syst. Rev. \textbf{40} (2006), no.~3,
  49--55.

\end{thebibliography}
\end{document}